\documentclass{pasa}

\usepackage{amsmath,amssymb,amsfonts}
\usepackage{algorithmic}
\usepackage{graphicx}
\usepackage{textcomp}
\usepackage[utf8]{inputenc}
\usepackage{xcolor}
\usepackage{makecell}
\usepackage{multirow}
\usepackage{hyperref}
\usepackage{soul}

\title[Deep Learning for Image Sequence Classification of Astronomical Events]{Deep Learning for Image Sequence Classification of Astronomical Events}

\author[Carrasco-Davis et al.]{Rodrigo Carrasco-Davis$^{1,7}$, Guillermo Cabrera-Vives$^{2,7}$, Francisco Förster$^{6,7}$, Pablo A. Estévez$^{1, 7}$, Pablo Huijse$^{3,7}$, Pavlos Protopapas$^{5}$, Ignacio Reyes$^{1, 7}$, Jorge Martínez-Palomera
$^{4,6,7}$ and Cristóbal Donoso$^{2}$   

\affil{$^1$Dept. of Electrical Engineering, Universidad de Chile}%
\affil{$^2$Dept. of Computer Science, Universidad de Concepción}
\affil{$^3$Informatics Institute, Universidad Austral de Chile}
\affil{$^4$Dept. of Astronomy, Universidad de Chile}
\affil{$^5$Institute for Applied Computational Science, Harvard University}
\affil{$^6$Center for Mathematical Modeling, Universidad de Chile}
\affil{$^7$Millennium Institute of Astrophysics, Chile}
}%

\jid{PASP}
\doi{10.1017/pas.\the\year.xxx}
\jyear{\the\year}

\usepackage{aas_macros}
\usepackage{hyperref} 
\hypersetup{colorlinks,citecolor=blue,linkcolor=blue,urlcolor=blue}

\hypersetup{draft}

\begin{document}

\begin{frontmatter}
\maketitle

\begin{abstract}

We propose a new sequential classification model for astronomical objects based on a recurrent convolutional neural network (RCNN) which uses sequences of images as inputs. This approach avoids the computation of light curves or difference images. This is the first time that sequences of images are used directly for the classification of variable objects in astronomy. 
The second contribution of this work is the image simulation process. We generate synthetic image sequences that take into account the instrumental and observing conditions, obtaining a realistic, unevenly sampled, and variable noise set of movies for each astronomical object. The simulated dataset is used to train our RCNN classifier. This approach allows us to generate datasets to train and test our RCNN model for different astronomical surveys and telescopes. Moreover, using a simulated dataset is faster and more adaptable to different surveys and classification tasks. We aim at building a simulated dataset whose distribution is close enough to the real dataset, so that a fine tuning could match the distributions and solve the domain adaptation problem. To test the RCNN classifier trained with the synthetic dataset, we used real-world data from the High cadence Transient Survey (HiTS) obtaining an average recall of 85\%, improved to 94\% after performing fine tuning with 10 real samples per class. We compare the results of our RCNN model with those of a light curve random forest classifier. The proposed RCNN with fine tuning has a similar performance on the HiTS dataset compared to the light curve random forest classifier, trained on an augmented training set with 10 real samples per class. The RCNN approach presents several advantages in an alert stream classification scenario, such as a reduction of the data pre-processing, faster online evaluation and easier performance improvement using a few real data samples. The results obtained encourage us to use the proposed method for \textit{astronomical alert brokers} systems that will process alert streams generated by new telescopes such as the Large Synoptic Survey Telescope.


\end{abstract}

\end{frontmatter}

\clearpage

\section{INTRODUCTION }
\label{sec:intro}


Astronomy is faced with the challenge of increasingly large streams of data produced by large survey telescopes. New telescopes, such as the Large Synoptic Survey Telescope \cite[LSST,][]{LSST1} and the Zwicky Transient Facility \cite[ZTF,][]{ZTF1} are designed to study variables and transients on wide areas of the sky. Variable stars, such as pulsating (e.g. RR Lyrae, Cepheids) or eclipsing stars; or transients, such as supernovae, are expected to be produced in large numbers. These objects have characteristic timescales from hours to months, and can be detected and characterized by repeatedly observing the same region of the sky. Obtaining these repeated images with large cameras will generate a very large volume of data. For example, it is estimated that the LSST will generate 30 TB of data per night to produce a complete image of the southern sky every 3 days. 

Some research areas in astronomy require the classification of a large number of objects: e.g. supernovae are used to estimate cosmological distances for studies about the expansion of the universe \cite[]{1998_supernovae,supernovae_2}, and variable stars such as RR Lyrae or Cepheids are needed to map the structure of the Milky Way and serve as \emph{cosmic distance ladders} \cite[]{distance1,milky2}.
In order to classify different astronomical objects in these large data streams we need to apply fast and accurate classification methods capable of managing large amounts of data in real-time. This problem will be addressed by systems called \textit{astronomical alert brokers} \cite[e.g. ALeRCE, LASAIR, ANTARES][]{ANTARES}, which are capable of receiving, processing, classifying and reporting relevant information about the alert streams generated by large survey telescopes in real time.


Traditional methods to classify variable astronomical objects are based on pre-processing a sequence of images (calibration) followed by feature extraction (measurement). One way to extract features from a sequence of images is doing photometry \cite[e.g.][]{Naylor}, which is the calculation of the total amount of light arriving from the source to the camera  as a function of time, generating a time series called a \emph{light curve}. In principle, a point--like source's light curve should contain all the relevant information about the source, but when the detection is spurious the information contained in the images becomes more relevant for the classification. Additionally, extragalactic sources such as supernovae tend to be near extended sources, i.e. galaxies, whereas galactic variable stars tend to be relatively isolated, information which is also contained in the pixels.

Obtaining the light curve reliably requires performing difference imaging first for certain sources (e.g. when the object occurs in a bright galaxy), which is the process of aligning, convolving and differencing pairs of images to show only those pixels which have changed from frame to frame \cite[e.g.][]{HiTS}.
Computing the difference image presents some problems, most of the time it is necessary to reduce the quality of one of the two images to subtract them correctly and it is also very sensitive to alignment errors between the frames.

Once the full light curves are computed, additional features can be extracted by manual design or automatic learning from the data. In the case of manual feature extraction, the scientist must design attributes that are expressive enough to contain relevant information for the classification, which usually requires a lot of effort and time \cite[e.g.][]{massive_lc_1,massive_lc_2,automatic_disc,using_machine,FATS,Protopapas2_probabilistic, Protopapas1_features}. Learning features directly from data is one way to avoid manual design and can be very useful to find informative attributes for classification \cite[]{IJCNN1, deep_hits,difference,gravitational_waves,exoplanets,lc_charnock,learnt,protopapas4_lstm}. However, even if representative features are obtained, if the data from the pre-processing step is not informative enough or contains errors from the procedure it would be difficult to obtain a good classification. Furthermore, in the case of an alert stream, for a new incoming sample of a light curve, it is desirable to update the feature value using the last point instead of using the entire light curve to avoid unnecessary computation and data retrieving \cite[]{2018NatAs...2..151N}.

Deep learning techniques are examples of data-driven solutions extracting features automatically that have proven to be successful in classification problems. Convolutional neural networks \cite[]{Original_cnn} have been applied to spatially correlated data such as images \cite[]{AlexNet,GoogleNet} and temporal correlated data such as audio \cite[]{audio1,audio2} among others. Recurrent neural networks, e.g. those containing Long Short Term Memory units \cite[LSTM, ][]{lstm1, lstm2}, have been applied to many natural language processing problems \cite[]{nlpbook} like translation \cite[]{translation1} and speech recognition \cite[]{lstm_speech}.

Recently, deep learning has been successfully applied to astronomical problems using convolutional neural networks, for example, for real/bogus separation  \cite[]{IJCNN1, deep_hits}, photometry computation \cite[]{cnn_japon}, calculation of an image comparable to the difference image \cite[]{difference}, gravitational wave detection \cite[]{gravitational_waves} and exoplanet detection \cite[]{exoplanets}. Recurrent neural networks have been used for light curve classification  \cite[]{lc_charnock,learnt,protopapas4_lstm, 2018NatAs...2..151N}. 


Recurrent convolutional neural networks are a special type of neural network where convolutional layers are combined with recurrent layers. Usually, a first stage of convolutional layers extract features from the raw data and generate high-level representations in deeper layers, then a second stage of recurrent layers uses the features yielded by the convolutional layers to learn time dependencies. Examples of applications are action recognition in videos \cite[]{video2,video1, video3} and speech recognition \cite[]{RCNN_speech}.

The first contribution of this work is the proposal of a model to classify variable astronomical objects based on a recurrent convolutional neural network (RCNN), which uses sequences of images directly as inputs. This way, we can estimate the class probability of a specific astronomical object by using the alert stream directly. With this approach, there is no need for recomputing features when a new observation arrives, we just feed the recurrent model with the incoming image. The information about previous images of the source is encoded in the network state. The computational cost of the proposed model scales linearly with the number of images within a sequence, encouraging us to use it when facing large data stream. Furthermore, we ensure that all the information available on the images is fed to the classifier without inducing errors by computing the light curve (e.g. in spurious sources) or the difference images (e.g. in badly convolved images). Our model consists of convolutional layers aimed at learning spatial correlations automatically from the images at each epoch, followed by a recurrent layer aimed at learning time dependencies between frames of an image sequence.
To the best of our knowledge, this is the first time that image sequences are used directly to classify astronomical objects. 


The second contribution of this work is the image simulation process. We generate synthetic image sequences that take into account the instrumental and observing conditions, obtaining a realistic, unevenly sampled, and variable noise set of movies for each astronomical object. The simulated dataset is used to train our RCNN classifier. This procedure is faster and more adaptable to different surveys and classification tasks, as compared to collecting real labeled data which is time-consuming and fixed to a specific survey. Furthermore, by randomizing simulations correctly we could generate a virtually infinite number of labeled samples, avoiding the problem of manually labeling a large number of real objects.  Simulating a dataset allows us, for example, to create data samples for telescopes that are still under construction such as the LSST. We can also tune the simulation parameters according to specific classification tasks and scientific objectives. 

Since the simulated data distribution may differ from those of real image sequences, we use a transfer learning technique called fine tuning \cite[]{2014arXiv1411.1792Y,6909618} to adapt our model trained over simulated images to solve the classification task on real images. By just using a few real labeled image sequences, we are able to improve substantially our RCNN model performance on the real dataset.

The structure of this article is the following: in Section~\ref{sec:sims} we describe the process of simulating synthetic images and in Section~\ref{sec:dl} we present a deep learning framework  with our proposed RCNN image classifier, and a light curve random forest classifier used for comparison purposes. In Section~\ref{sec:results} we present the classification results obtained using the RCNN model, and compare it with a light curve random forest classifier. In addition we show the results of performing fine tuning with a few real images for both classifiers. In Section~\ref{sec:discussion} we discuss the main implications of this work and in Section~\ref{sec:conclusions} we draw the main conclusions and describe future steps.

\section{Data Simulation} \label{sec:sims}

We built a simulated dataset of labeled image sequences, using the following procedure. First, we gather all the information required to mimic realistic observing conditions which correspond to instrument specifications, observation dates, exposure times and atmospheric conditions. These parameters are specified in Section \ref{sec:simulation_params}. In this work we simulate observing conditions for the HiTS survey 2015 on band $g$ \cite[]{HiTS, 2018NatAs.tmp..122F, 2018arXiv180900763M}. Next, we simulate light curves based on physical and empirical models, and sample from them using the observation dates. The instrument specifications, exposure times and atmospheric conditions are used to generate an image for each point of a light curve, and finally noise is added to each image. This way, we produce an irregularly sampled movie of 21x21 pixels for each astronomical object. The simulated image sequence dataset is used to train the proposed RCNN model, which is explained in Section~\ref{sec:dl}. In the next sections we explain the image simulation process in detail. 


\subsection{Synthetic data simulation parameters}
\label{sec:simulation_params}

\begin{table}[b!]
\centering
\caption{Image Simulation Parameters: Camera parameters are constant for a given instrument, but exposure parameters vary in time.}
\small
\begin{tabular}{|c|c|}
\hline
\multicolumn{2}{|c|}{\textbf{Camera parameters}}                      \\ \hline
Gain   [e-/ADU]                   & Read Noise     [e-]                \\ \hline
Saturation    [ADU]            & Pixel Scale          [arcsec/pixel]          \\ \hline
\multicolumn{2}{|c|}{\textbf{Exposure parameters}} \\ \hline
Date [MJD] & Seeing [pixels]                         \\ \hline
Airmass & Sky brightness [ADU]                 \\ \hline
Zero Point [mag] & Filter [$g, r, i$ or $z$] \\ \hline
Exposure Time [sec]  & Limiting magnitude [mag] \\ \hline
\end{tabular}

\label{table:sim_params}
\end{table}

The simulation process is adjustable to different observing conditions, so our model can be trained and applied to different instruments. To apply the RCNN classifier to a different survey, we should gather representative parameters, simulate images and train a new model on the simulated dataset. Ideally the distribution of the simulated data should be equal to the distribution of the real data. In practice, this is not always possible, so a domain adaptation problem arises where the features and tasks are the same but the distributions are different. Our aim is to build a simulated dataset whose distribution is close enough to the real one, so that a fine tuning phase with a few real samples could match the distributions and solve the domain adaptation problem.

The observing condition parameters are composed of camera and exposure parameters. A camera has a unique list of parameters describing the conversion from photons to digital units. An exposure has a unique list of parameters which describe the time and duration of the exposure, as well as the relevant atmospheric conditions during exposure. These parameters are summarized in Table \ref{table:sim_params}. Herein, we simulate data for the HiTS survey, which consists of 50 fields observed by DECam \cite[]{DECAM} with 62 CCD cameras per field, and between 25 and 30 observations per field. In this work, we use empirical observing conditions which are sampled from real observations from HiTS. The typical observing conditions are described in \cite{HiTS}.

The simulated images are produced assuming a given point spread function (PSF), which is sampled from a collection of empirical PSFs, an efficiency of conversion from physical units to analog digital units given by the camera and exposure parameters, and the sky level given in the exposure parameters. More details are given in Section~\ref{sec:image_sim}.

\subsection{Light curve simulation}

\begin{figure}[!b]
\centerline{ \includegraphics[width=0.48\textwidth]{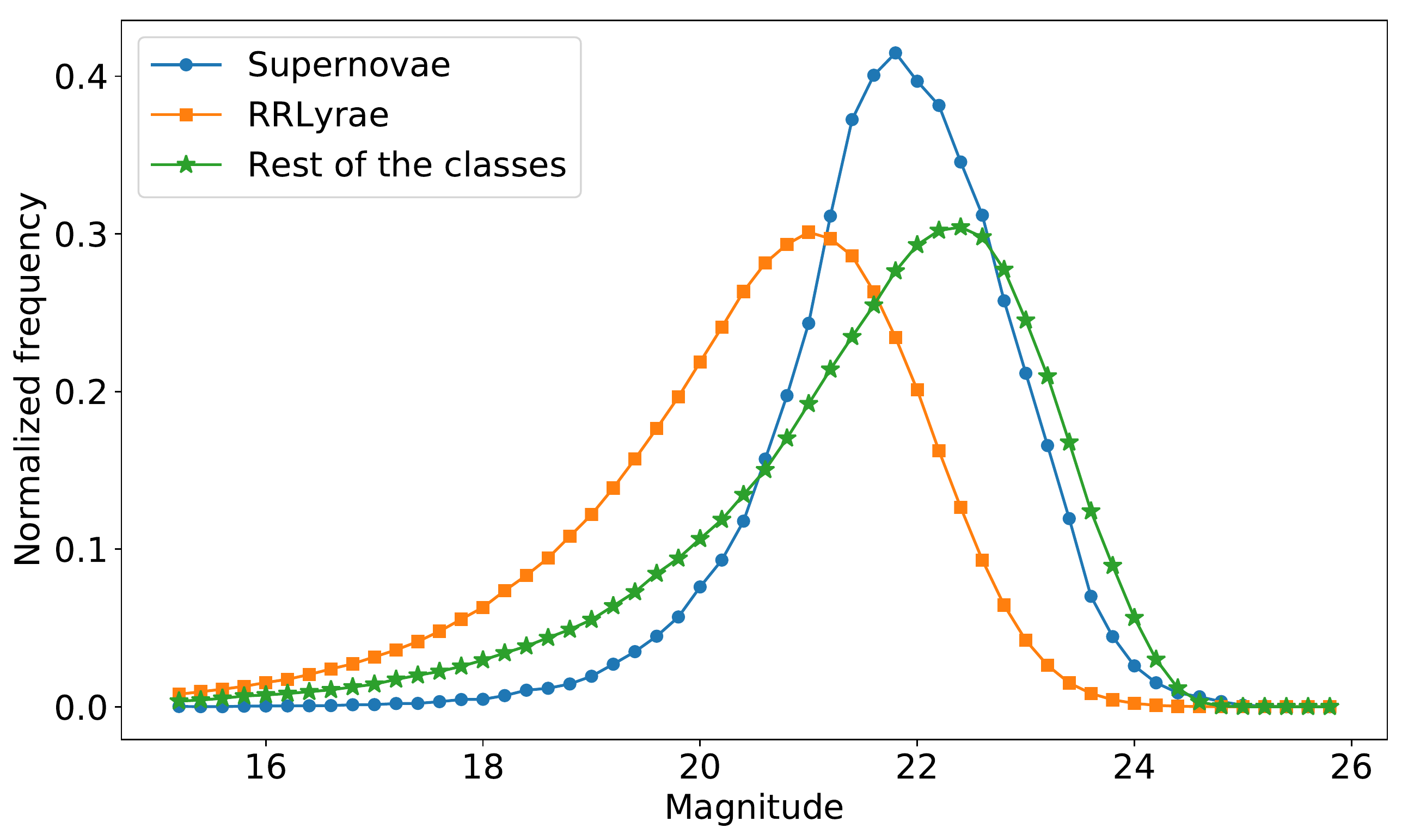}}
\caption{Magnitude density distribution of the simulated data.}
\label{fig:mag_distr}
\end{figure}

We use seven classes of astronomical objects (see Table~\ref{table:Classes}): two non--variable (non--variable stars and galaxies) and five variable or transient (RR Lyrae, Cepheids, eclipsing binaries, supernovae and asteroids). 
Variable sources are simulated in two steps: 1) sampling from either a physical model or empirical data, and 2) adjusting their brightness by sampling from certain magnitude distributions as explained below.
In order to sample each type of light curve for a given observation date we used different interpolation methods.

\begin{table*}[t!]
\centering
\caption{Class description, astronomical sources simulated in this work.}
\label{table:Classes}
\begin{tabular}{|c|c|}
\hline
\textbf{Astronomical Object} & \textbf{Generation model} \\ \hline
Supernovae                   & \makecell{Simulations based on physical models of SNe II from \cite[]{Moriya} \\ and SNe Ia spectrophotometric templates from  \cite[]{2007ApJ...663.1187H} }                            \\ \hline
RR Lyrae                     &          \makecell{483 light curve templates, sampling \\ a random phase and average magnitude \cite[]{RRLyrae}}                    \\ \hline
Cepheids                     & \makecell{600 real cepheids light curve \cite{Cepheids} fitted \\ with a Gaussian process for interpolation \cite[]{GaussianProcess}}                               \\ \hline
Eclipsing Binaries           &     \makecell{375 Eclipsing binaries templates from \\ CatSim\footnotemark, part of the LSST simulation tools}                     \\ \hline
Non-Variable objets          &                      \makecell{Constant brightness value \\ for each time of observation}     \\ \hline
Galaxies                     &       \makecell{Exponential and De Vaucouleur's luminosity profile \\ using parameters from SDSS galaxy catalog \cite[]{SDSS}}                   \\ \hline
Asteroids                    &   \makecell{Simulated as a bright source \\ in just a single time of observation} \\ \hline  
\end{tabular}
\end{table*}
\footnotetext{\url{https://www.lsst.org/scientists/simulations/catsim}}

We start by sampling a light curve either from a physical model or from empirical data. Table~\ref{table:Classes} shows the sources of the light curves we sampled from. 
Supernova redshifts and light curves are obtained from simulations which take into account cosmology and supernova rates, the telescope parameters, and physical models for SNe II from \cite{Moriya} and spectrophotometric templates for SNe Ia from \cite{2007ApJ...663.1187H}.
 RRLyrae, Cepheids, and eclipsing binaries were sampled from real data from \cite{RRLyrae},  \cite{Cepheids}, and using the LSST Catalog Simulation database (\url{https://www.lsst.org/scientists/simulations/catsim}, CatSim) respectively. Light curves for non--variable objects were simply simulated as a constant light curve, and asteroids as a single peak. Galaxy simulations are explained in Section~\ref{sec:image_sim}.

We sample light curves using the empirical exposure parameters from HiTS and scale them to follow a magnitude distribution that reproduces the HiTS observations. Magnitudes for non--variable sources, eclipsing binaries, Cepheids, and asteroids are sampled from the green curve (rest of the classes) in Figure~\ref{fig:mag_distr}. This distribution was obtained by fitting an exponential function to the distribution of stars in HiTS. A constraint was added to smooth the decay at large magnitudes in order to follow the supernovae magnitude distribution. This was performed by multiplying the exponential density distribution by a cutoff function $f(m, m_\mathrm{cutoff}) = 1 - \mathrm{erf}(m - m_\mathrm{cutoff})/2$, where erf is the error function and $m_{\rm cutoff}$ is the value where $f(m, m_\mathrm{cutoff}) = 0.5$. The supernovae magnitude distribution decay at large magnitudes can be mimicked by using $m_{\rm cutoff} = 22.8$ for all the classes, except for RR Lyrae and galaxies. For RR Lyrae, $m_{\rm cutoff} = 21.5$ was chosen to make the distribution with magnitude boundaries based on \cite{medina2018}. 

\begin{figure}[!b]
\centerline{ \includegraphics[width=0.54\textwidth]{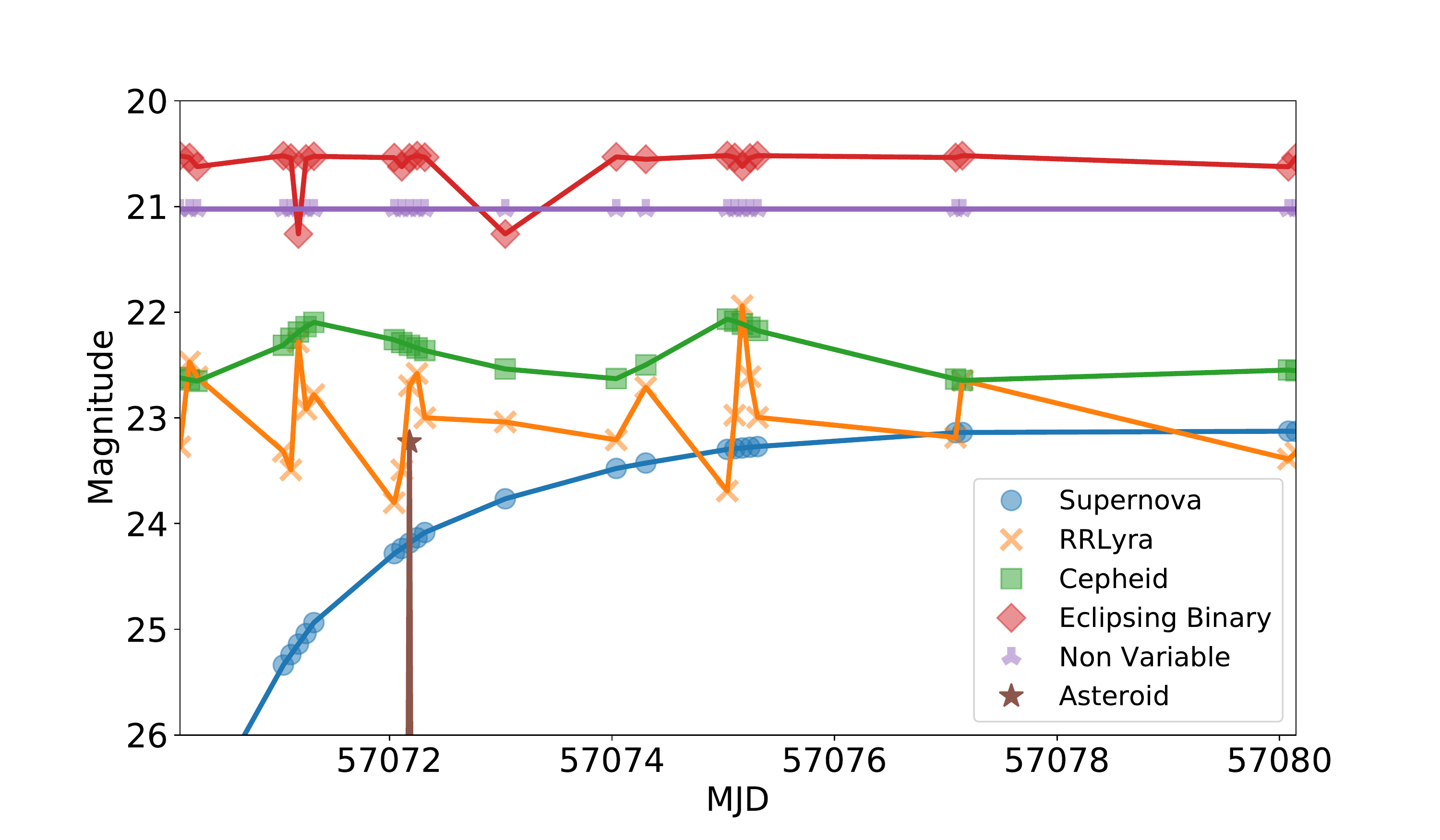}}
\caption{Light curve examples for six classes of astronomical objects in the dataset. Galaxies are simulated by using the method described in section \ref{sec:image_sim}}.
\label{fig:lightcurve_samples}
\end{figure}

At this point, simulated light curves have no noise. Errors of flux in real light curves are estimated from statistical assumptions about measurement noise within the image, so the noisy versions of simulated light curves are recovered from noisy simulated images in later steps. Fig. \ref{fig:lightcurve_samples} shows examples of light curves for each category (except for the galaxy class).

\subsection{Image simulation using light curves}
\label{sec:image_sim}

\begin{figure}[t!]
\centerline{
\includegraphics[width=0.55\textwidth]{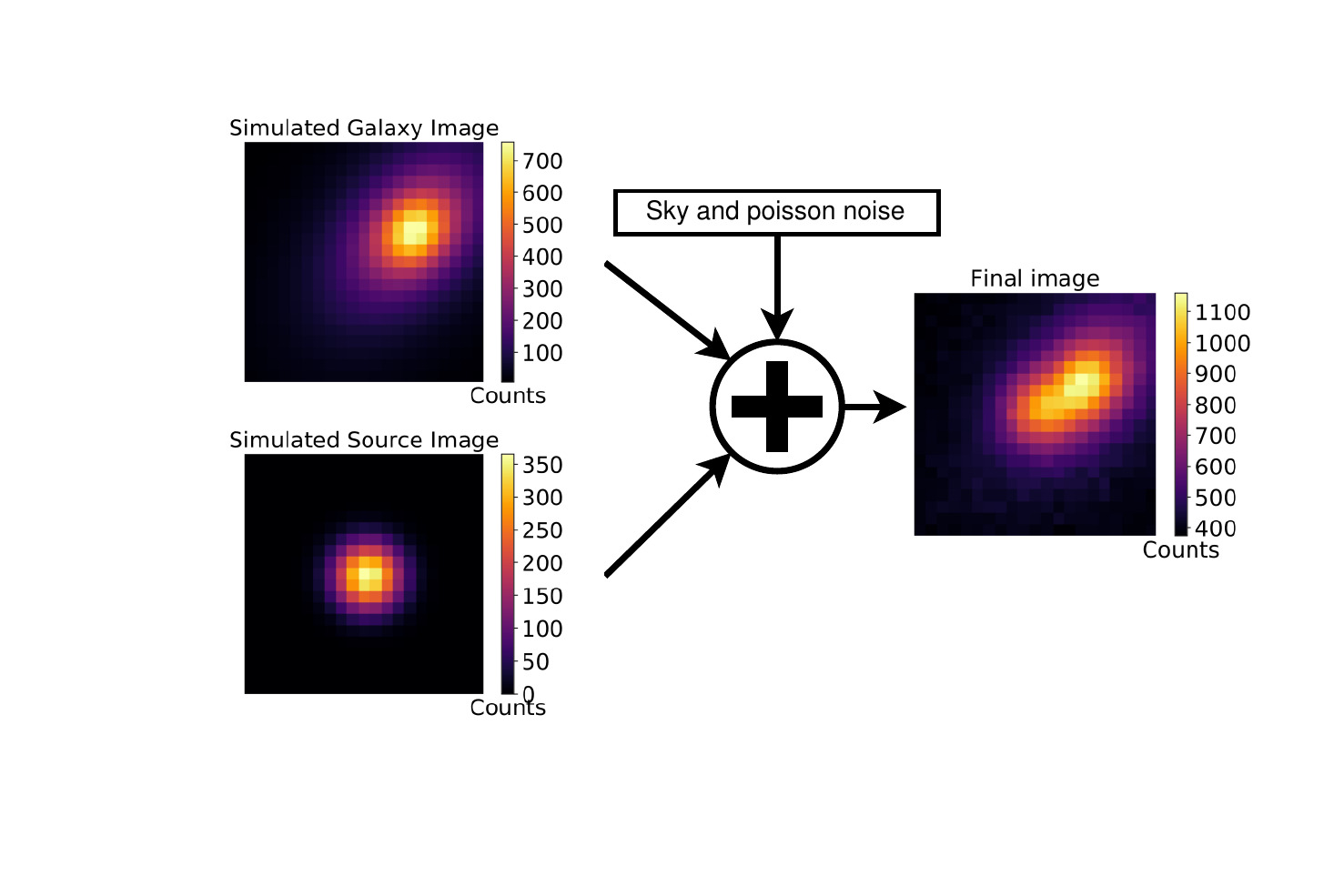}
\vspace{-1cm}}
\caption{Summary of the image simulation process. The light coming from a source is spread in the Source Image. A simulated host galaxy is added to the source image. The sky brightness is added as a constant value to all the pixels and Poisson noise is sampled with variance equal to the number of photons in each pixel along with CCD readout noise.}
\label{fig:image_sim}
\end{figure}

Having simulated light curves sampled with the corresponding cadence, we used the zero point values $Z_{p}(t)$ to convert each point of the light curve from magnitudes to ADU (analog-to-digital) using: 
\begin{equation}
\label{eq:counts_to_mag}
m(t)=Z_{p}(t)-2.5\log{\left (\frac{\text{ADUs}(t)}{T(t)}\right )}, 
\end{equation}
where $t$ is the observation time and  $T(t)$ is the exposure time for an image at time $t$. Usually, there are other terms in this conversion associated with airmass and color, but these $Z_{p}$'s were computed using PanSTARRS1
\cite[][]{2016arXiv161205560C} to fit the resulting magnitude of known sources. For each light curve we choose a random CCD array and use its exposure parameters at different epochs. 
Then, for each point in ADU units of the light curve, a point spread function (PSF) $p_{t}(x, y)$ is used to generate a source image, where $x, y$ are pixel coordinates and $\sum_{x,y} p_{t}(x, y)=1$. We generate a source image (see the example shown in Figure \ref{fig:image_sim}) by creating an empty image of $21\times 21$ and adding $
\text{ADUs}(t) \cdot p_{t}(x-x_{0}, y-y_{0})$  where $x_{0}, y_{0}$ is the center of the source in the image, sampled from a uniform distribution within the single image central pixel to simulate random centering errors. The PSF $p_{t}$ is estimated by averaging real source images 
from the HiTS survey and computing its FWHM by fitting a 2D-Gaussian function as an estimation of size. We used PSFs estimations with different sizes, but for each observation time $t$ we match a single $p_{t}$ with the current FWHM(t). Dates with FWHMs larger than 2'' were not used. 
 A random rotation and mirroring is applied to make the classifier invariant to rotations of the PSF.

Some sources may have a host galaxy, so we simulated them by using 
exponential and De Vaucouleurs profiles with parameters obtained from the Sloan Digital Sky Survey \cite[SDSS,][]{SDSS}, including the following (if many bands are used, these quantities are per band): radii, ellipticities, proportion between the two profiles, 
and the luminosity of the galaxy in magnitudes. The magnitudes were converted to ADUs and distributed using the exponential and De Vaucouleurs profiles. These profiles have a spike in the center, concentrating most of the flux in the central pixel. 
In order to avoid this issue, we sampled 20 uniform random positions in the range of the central pixel, computed the bulge profile and averaged them to distribute the flux correctly across the image. We finally convolved this image with $p_{t}(x, y)$ generating a galaxy image $\text{IM}_{\text{gal}}$. In order to simulate a SN in a host galaxy,
we sampled the position from a distribution following the exponential profile.

The last step for image simulation is producing a joint image by adding up the PSF-like image, the galaxy, and the sky brightness $\text{Sky}(t)$ for time $t$.
Then, we convert ADU pixels to electrons $e-$ multiplying by the corresponding Gain of the camera, in order to apply independent Poisson noise to each pixel and Gaussian readout noise. Finally, the image is converted back to ADUs. The resulting image sequence for each object is unevenly sampled. Also, the noise on each stamp is variable and consistent with a realistic observation since it depends on the observation conditions such as the PSF size, sky brightness, zero point estimation, presence of a galaxy, along with the CCD readout noise.
An example of a simulated SN with a host galaxy image is shown in Figure \ref{fig:image_sim}.

Host galaxies were added on to 50\% of the supernovae objects and 5\% of the rest of the classes. Although these proportions are artificial, they are inspired on prior knowledge about the abundance of supernovae occurrence on host galaxies and the low probability of finding other classes near a host galaxy. As mentioned before, the galaxy class is a simulated image of a host galaxy with varying exposure parameters.

Examples showing both real image sequences from HiTS and simulated sequences for non--variable objects can be found in Appendix A. 

\section{Deep learning framework} \label{sec:dl}

Our goal is to discriminate among the seven classes shown in Table \ref{table:Classes}. We simulated 686,000 objects for the training set, 85,750 for the validation set and 85,750 for the test set. Each set has a balanced number of objects per class. 

\label{classifier_model}

\begin{table}[!b]
\centering
\footnotesize
\caption{Recurrent Convolutional Neural Network architecture.}
\label{table:architecture}
\begin{tabular}{|c|c|c|}
\hline
\textbf{Layer}  & \textbf{Layer Parameters}           & \textbf{Output Dim}      \\ \hline
Input Layer   & $21\times 21 \times n_{w}^{\mathrm{a}}$       & $21 \times 21 \times n_{w}$  \\ \hline
BRN  & $n_{w}$ (mean and std)             & $21 \times 21 \times n_{w}$  \\ \hline
Conv + BRN       & $3 \times 3 \times 64$, $64$ & $21 \times 21 \times 64$ \\ \hline
Conv + BRN   & $3 \times 3 \times 64$, $64$ & $21 \times 21 \times 64$ \\ \hline
Conv + BRN   & $3 \times 3 \times 64$, $64$ & $21 \times 21 \times 64$ \\ \hline
Max pooling      & $2 \times 2$, stride 2       & $11 \times 11 \times 64$ \\ \hline
Conv + BRN   & $3 \times 3 \times 64$, $64$ & $11 \times 11 \times 64$ \\ \hline
Conv + BRN   & $3 \times 3 \times 64$, $64$ & $11 \times 11 \times 64$ \\ \hline
Conv + BRN   & $3 \times 3 \times 64$, $64$ & $11 \times 11 \times 64$ \\ \hline
Max pooling      & $2 \times 2$, stride 2       & $6 \times 6 \times 64$   \\ \hline
\makecell{Fully connected \\ (with dropout)} & $2304 \times 1024$           & $ 1024 $                 \\ \hline
LSTM            & \makecell{ 1024 + $\Delta t$ of samples \\ $512$ units}          & $512$                    \\ \hline
Output softmax & $512 \times 7$              & $7$ (n° classes)                     \\ \hline

\end{tabular}
\vspace{0.05cm}
\footnotesize 

$^{\mathrm{a}}n_{w}$ is the number of images stacked in the input tensor, BRN stands for Batch Renormalization 
\end{table}

\begin{figure*}[!t]
\centerline{\includegraphics[width=0.95\textwidth]{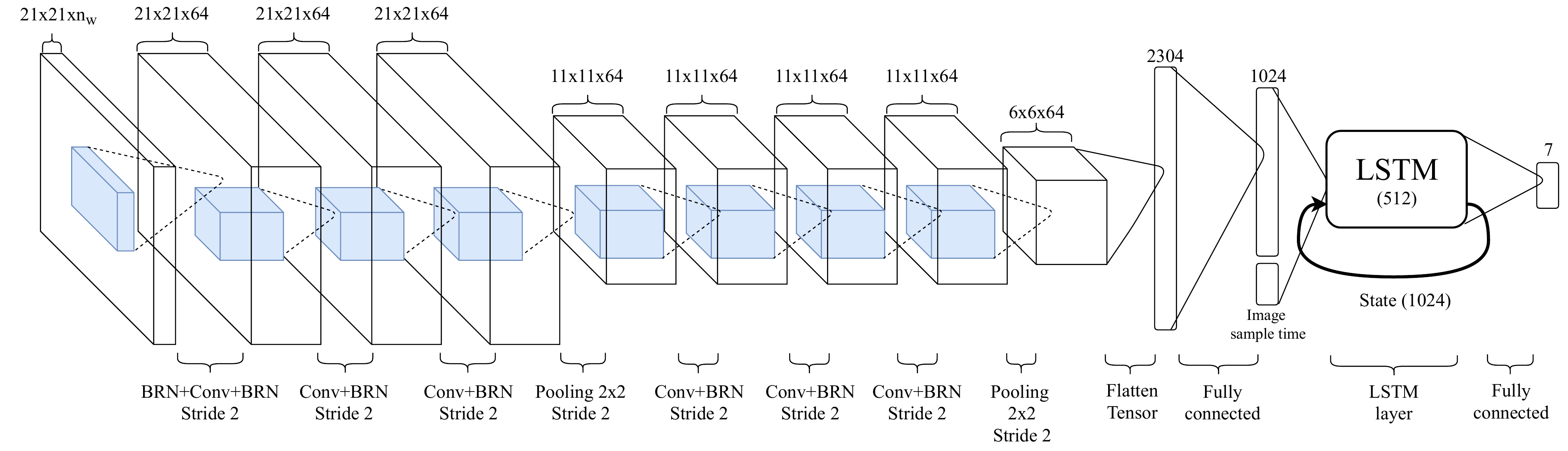}}
\caption{RCNN architecture. An input tensor with shape $(21, 21, n_{w})$ is shown at the left of the image. Every layer is described at the bottom and the shape of the data is indicated at the top.
}
\label{fig:architecture}
\end{figure*}

We propose a new model to classify astronomical objects based on a recurrent convolutional neural network (RCNN), which uses sequence of images as input. Convolutional layers are able to automatically learn the spatial correlation between pixels in the input image and extract high-level features, which are used by the recurrent layer to learn time dependencies among images sampled at irregular times. The components of the RCNN architecture (see Fig. \ref{fig:architecture}) are the following:

\paragraph*{Fully connected layer:} Usually placed at the top of the architecture. It takes high-level representations obtained by lower layers and applies the transformation $Y = f(WX+b)$, where $X$ is the input of the layer, $W$ are the weights, b is the bias and $f$ is a non-linear activation function. Typical choices for $f$ are the rectifying linear unit ReLU \cite[]{ReLU}, hyperbolic tangent or logistic sigmoid.

\paragraph*{Convolutional layers:} These layers apply convolutional operations to their inputs. The most popular one is the convolution over images done by filters, which are adjusted through the training process \cite[]{Original_cnn,AlexNet,GoogleNet}. The convolution $Y$ of an image $X$ with a filter $W$ plus a bias $b$ is expressed as:

\begin{equation}
Y_{i,j,k} =  \sum_{m,n,p}X_{i-m, j-n, p}\cdot W_{m,n,p,k} + b_{k},
\end{equation}
where $X$ is a 3D tensor, $i, j$ are the image coordinates, the indexes $m, n$ run over a sub-matrix of the image and the index $p$ runs over the depth or the number of channels. $W$ has one more dimension $k$ representing the output channels. Right after the convolution operation, a non-linear activation function is used, the same way as in the fully connected layer.

\paragraph*{Pooling layers:} These layers are applied to reduce the dimensionality of the representation  through the network and act as a regularizer. In this work, we used max pooling operations that return the maximum value within a sub-matrix of the image,
$ Y_{i,j} =  \text{max}\left ( X_{i:(i+n), j:(j+m)} \right ) $ where $i, j$ are the image coordinates and $n, m$ are the coordinates of the sub-matrix.

\paragraph*{Batch Renormalization:} It is an extension of Batch Normalization \cite[][]{BatchNorm} used to standardize the values of the variables by shifting and scaling them. Some of the effects of batch normalization are faster training speed and better regularization. Batch renormalization helps to match the normalization process for training and inference while preserving advantages from regular batch normalization \cite[][]{2017arXiv170203275I}. The batch renormalization operation is during training is:
\begin{equation}
Y = \left ( \frac{X - \mu_{\text{batch}}}{\sigma_{\text{batch}}} \cdot r + d \right ) \gamma + \beta,
\end{equation}   where $\mu_{\text{batch}}$ is the mean and $\sigma_{\text{batch}}$ is the standard deviation for the current batch, and $\gamma$ and $\beta$ are trainable parameters. The renormalization extension corresponds to $r=\text{clip}_{[1/r_{\max},r_{\max}]} \left ( \frac{\sigma_{\text{batch}}}{\sigma_{d}} \right )$ and $d=\text{clip}_{[-d_{\max}, d_{\max}]} \left ( \frac{\mu_{\text{batch}}-\mu_{d}}{\sigma_{d}} \right )$, where $\mu_{d}$ and $\sigma_{d}$ are the mean and standard deviation of the data computed using a moving average $\mu_{d} = \alpha \mu_{d} + (1-\alpha)  \mu_{\text{batch}}$ with $\alpha$ a momentum parameter, $r$ and $d$ are taken as constant when computing the gradients and setting $r=1$ and $d=0$ we recover the original batch normalization. In the case of convolutional layer outputs, this operation is applied to each channel independently.

\paragraph*{Recurrent Layers and LSTM:} It is the part of the network that learns time dependencies in a sequence of inputs. There are many types of recurrent layers and networks \cite[][]{Recurrent} but most of them have a feedback connection to the input from previous time steps or a state (or both), where a state is an arbitrary representation of a memory of previous inputs. In particular, the recurrent model used in this work is the Long Short Term Memory \cite[LSTM,][]{lstm1,lstm2}. The main characteristic of LSTM is the use of gates that control the content of the state in order to learn longer time dependencies than conventional recurrent networks. LSTM has three main gates: the forget gate removes part of the state using the information from current input and previous output, the input gate updates the state, and the output gate combines input, state, and previous output.

\subsection{Sequence input to the model}
\label{inputs_to_model}
Before inputting an image to the classifier we perform a pre-processing step that consists in subtracting the sky in counts to the image as a constant value. Then, we multiply each pixel by a factor that ensures the same number of counts for a given magnitude, making the conversion in equation \ref{eq:counts_to_mag} invariant to the zero point $zp$. The operation is the following:
\begin{equation}
\label{eq:preprocessing}
\text{IM}_{\text{proc}}(x, y) = (\text{IM}_{\text{orig}}(x,y) - \text{sky})\cdot 10^{\frac{zp_{\text{ref}}-zp}{2.5}}, 
\end{equation}
where $\text{IM}_{\text{proc}}(x, y)$ is the resulting preprocessed image, $\text{IM}_{\text{orig}}(x,y)$ is the original image, and $zp_{\text{ref}}$ is a reference zero point, chosen as the zero point of the first exposure for each field. We use $\text{IM}_{\text{proc}}(x, y)$ to build the inputs to the model.

In the case of supernovae and asteroids, we considered objects where the source is detectable in at least one of the images within a sequence i.e., at least one point of the original light curve in magnitude must be above the limit of magnitude. The first point when this happens is the ``first alert'' triggered by the rise of flux in time. 

Because we are interested in classifying supernovae in the early stages of their explosion, once an alert is triggered (supernova explosion or asteroid appearance) at time $t_{i}$, we query the five images previous to the alert and create a stack of $n_{w}$ consecutive images using $t_{i-5}$ as the first image of the stack. In this work we used $n_{w} = 3$, so if the alert occurs at time $t_{i}$, then the input images for the first stack are at time $(t_{i-5}, t_{i-4}, t_{i-3})$, then next time step input will be at $(t_{i-4}, t_{i-3}, t_{i-2})$, then $(t_{i-3}, t_{i-2}, t_{i-1})$ and so on. Therefore the input to the model is a stack of $n_{w}$ consecutive images forming an input tensor of shape $(21 \times 21 \times n_{w})$. 

We define $N_{d}$ as the number of available dates, which corresponds to the number of points between five images before the first detection and last exposure on the respective field. $N_{d}$ depends on the first detection date for supernovae or asteroids. We use the same $N_{\mathrm{d}}$ dates to build the input sequence of the rest of classes (RR Lyrae, Cepheids, eclipsing binaries, non variables and galaxies), in order to have variable size and observation conditions for every class in the dataset. We truncated the maximum number of available dates to $N_{d} = 22$ for every object to evaluate the model. 

We use $n_{w}>1$ images stacked at the input and not a single image because convolutional layers can learn part of the short timescale dependencies between these $n_{w}$ consecutive images, letting the recurrent layer learn about longer timescale dependencies. We also compute the difference in sampling time between the first image of the sequence and the rest of the images, and feed the model with this irregular sampling for each image. 

\subsection{Image sequence classifier architecture}

\label{architecture}

Our model uses high-level representations of the image obtained by convolutional layers as inputs to the recurrent layer. This way, we can add information to the memory of the classifier while the images are received by changing one image at a time. The LSTM units in the recurrent layer contain memory cells that store learned knowledge from past input images.

As mentioned above, the input is a 3D tensor of size $(21, 21, n_{w})$, made by $n_{w}$ consecutive $21 \times 21$ images. We first apply a batch renormalization layer, followed by a convolutional layer to increase the number of channels from $n_{w}$ to 64 and an extra batch renormalization layer. We use 64 filters and a stride of 1 on each convolutional layer with ReLU as activation function. Also at the output of each convolutional layer a batch renormalization layer is implemented for each channel, a pool layer after the first three convolutional layer + batch renormalization, followed by three convolutional layers + batch renormalization and a final pool layer. 

The output of the pool layer is flattened to a vector of size $6 \times 6 \times 64 = 2304$,  which is the input to the first fully connected layer with 1024 hidden units. The time difference between the day of observation of the $n_{w}$ images and the first image of the entire sequence ($n_{w}$ size vector) is added to the input of the LSTM layer which has 512 units. The initial state of the LSTM is an array filled with zeros and the state is updated for every input tensor with $n_{w}$ stacked images. Finally, the LSTM output is passed through a fully connected layer with softmax activation functions. The details of each layer are shown in Table \ref{table:architecture}. Fig. \ref{fig:architecture} shows an illustration of the RCNN architecture.

\subsection{Light curve random forest classifier}

For comparison purposes, we designed a light curve classifier using feature extraction and a random forest (RF) classifier. The light curves in ADUs were extracted from the same set of simulated stamp images given to the RCNN using optimal photometry \cite[]{Naylor}. We computed all the features available in FATS \cite[]{FATS} for each light curve, except for the ones associated with color, \textit{Color, Eta\_color, Q31\_color, StetsonJ, StetsonL} since the HiTS survey uses mostly the g band. In order to evaluate the accuracy of the light curve RF classifier on the real dataset as a function of time, we recomputed the FATS features for each new point added to the light curve. FATS includes the estimation of the period of the light curve and other features from Lomb Scargle Periodogram \cite[]{1976Ap&SS..39..447L}, which are known to be important for variable star classification. The light curves are almost perfectly derived from the images, since we use the same information used to simulate them such as the PSF, presence of a galaxy and the background value to extract the light curve correctly. In practice, these parameters are estimated from the images. Finally, we trained a RF classifier \cite[]{Breiman2001}, to discriminate among the seven classes. Feature based random forest classifiers are commonly used in astronomy \cite[]{HiTS, 0004-637X-777-2-83,2014A&A...566A..43K,1538-3881-153-4-170}.

\subsection{Training Process}
\label{training}
Recurrent neural networks are trained using variants of gradient descent, by using backpropagation through time. Since features are extracted from the images using convolutional layers and are fed to the recurrent layer, the gradients through time are also used to adjust the parameters of the convolutional layers. 
For our RCNN model, we used cross-entropy as loss function to compare the outputs of the model with the labels. The total loss of a single example is $\text{loss} = \sum_{t}^{N_{d}}\text{loss}(t)$ where $\text{loss}(t)$ is the cross-entropy at time step $t$. The training algorithm is AMSGrad \cite[]{j.2018on} which is an adaptive learning rate algorithm. The batch size was 256 sequences of simulated images and we run 30,000 iterations presenting a single batch per iteration to the image sequence classifier, using $5\cdot10^{-4}$ as learning rate. The final model was chosen by selecting the one that had the lower loss in the validation set during the training process. We used graphic processor units (GPUs) to train the models. Each model takes approximately 8 hours to complete 30,000 iterations, equivalent to 9.5 epochs, in a GeForce GTX 1080 Ti. For the fine-tuning phase, after training RCNN with the simulated dataset, we run 1000 iterations with 10 image sequences per class randomly selected from the real dataset.

For the light curve classifier, we trained a RF using the light curves extracted from the simulated images, on which 58 FATS features were computed. We run a grid search for the best max depth of each decision tree and the number of them. The best results in the validation set of the simulated dataset were obtained with 25 estimators and a depth of 15. In order to make a fair comparison with the fine tuned version of the RCNN, we also trained a RF adding 10 samples of light curves per class from the real dataset to the simulated training set, resulting in an augmented training set. Each real sample added to the augmented training set was copied 100 times on the training set, which is comparable to the number of times that the RCNN observed every additional real example during the 1000 extra iterations.

\section{Results} \label{sec:results}

\begin{figure}[!b]
\centerline{\includegraphics[width=0.55\textwidth]{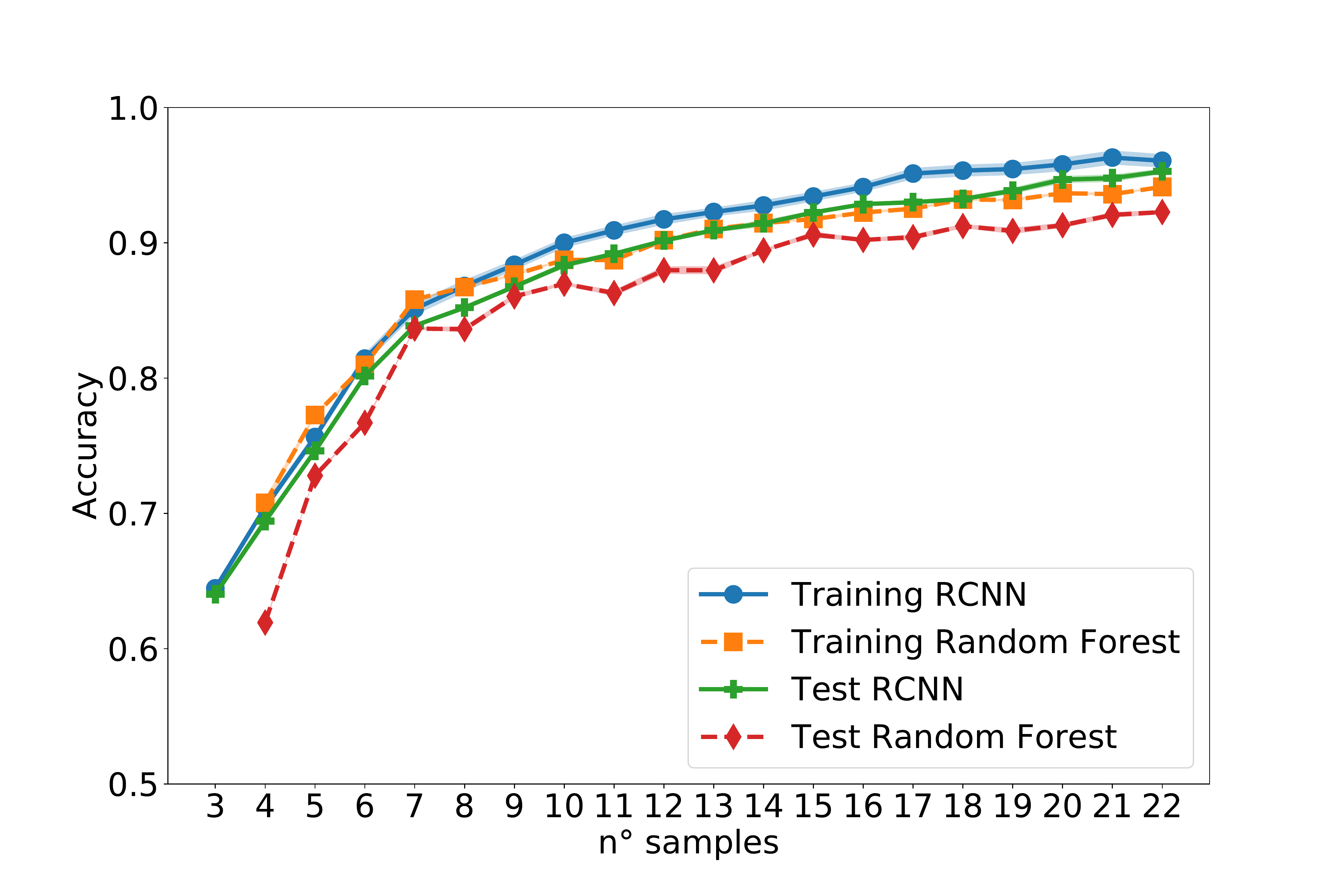}}
\caption{Accuracy on the simulated dataset as a function of the number of samples of the sequence available to the classifiers. The accuracy obtained at the last point (after 22 samples) for each curve is: Training RCNN: 0.961 $\pm$ 0.005, Training Random Forest: 0.941 $\pm$ 0.001, Test RCNN: 0.953 $\pm$ 0.002 and Test Random Forest: 0.923 $\pm$ 0.002.}
\label{fig:sim_image_performance}
\end{figure}

\begin{figure}[!b]
\centerline{\includegraphics[width=0.55\textwidth]{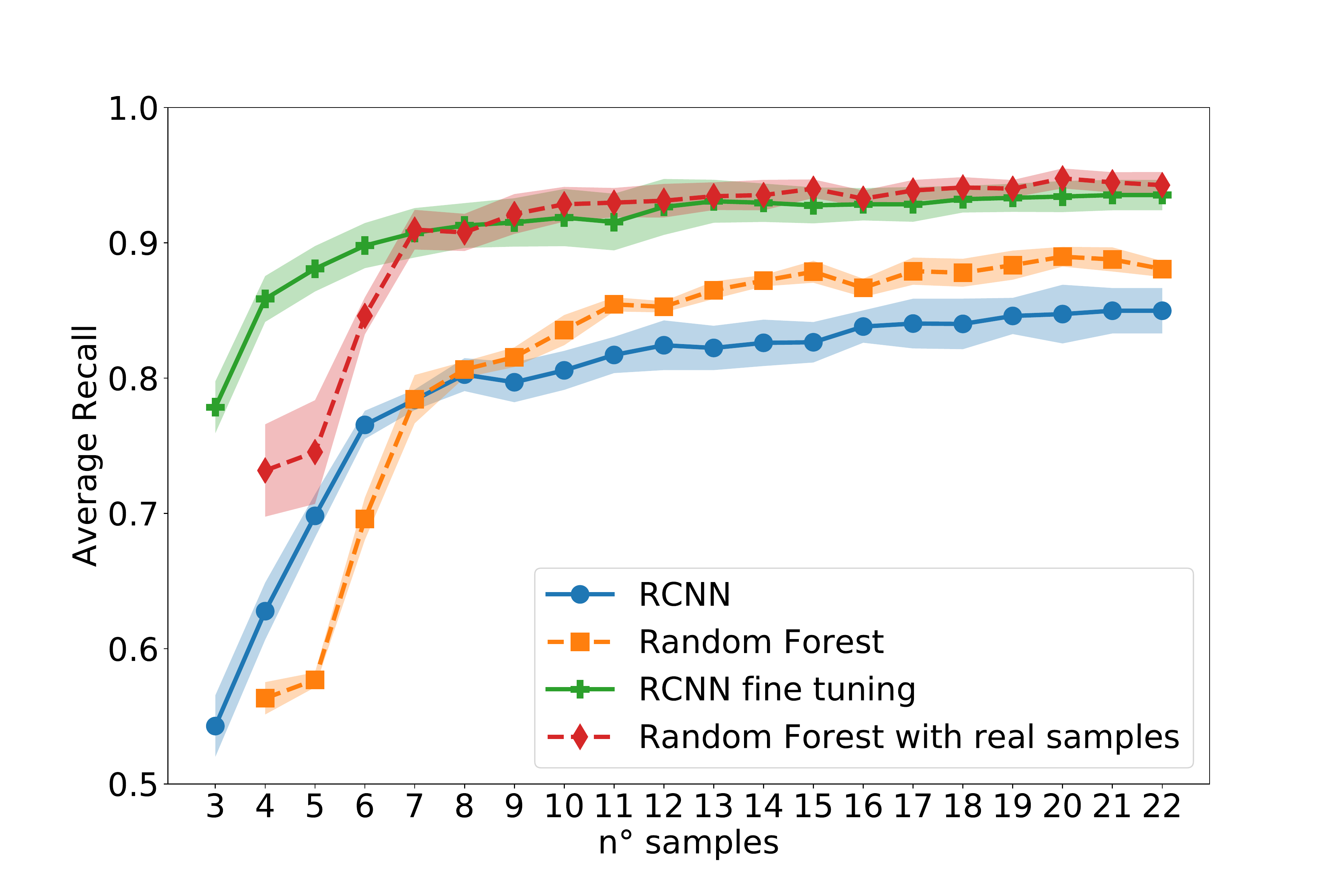}}
\caption{Average recall on the real dataset without and with fine tuning as a function of the number of samples of the sequence available to the classifiers. The average recall obtained at the last point (after 22 samples) for each curve is: RCNN: 0.85 $\pm$ 0.02, Random Forest: 0.88 $\pm$ 0.01, RCNN with fine tuning: 0.94 $\pm$ 0.01 and Random Forest trained with augmented training set: 0.94 $\pm$ 0.01. For the last to cases, the evaluation set does not include samples used to fine tune the model.}
\label{fig:real_image_performance}
\end{figure}

\begin{figure*}[!t]
\centerline{\includegraphics[width=1.2\textwidth]{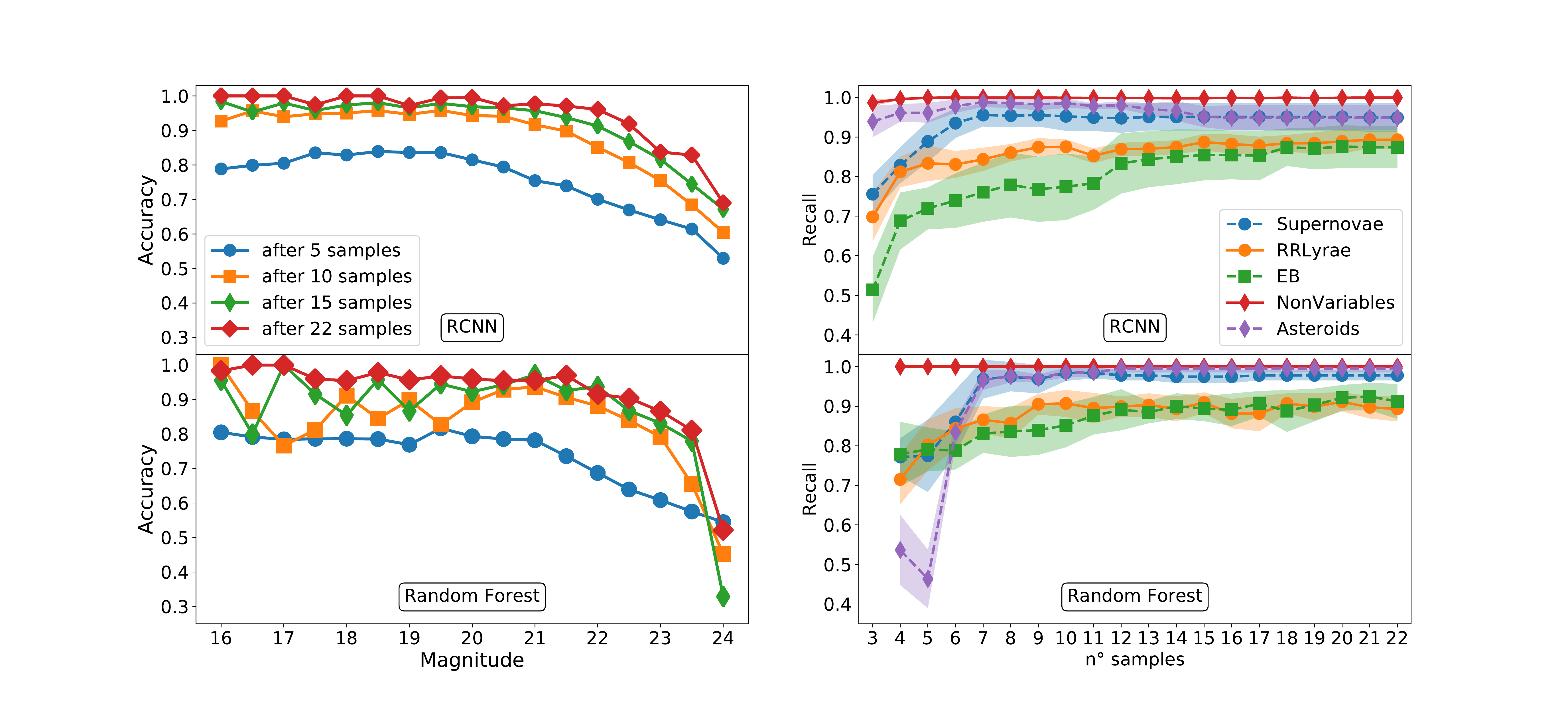}}
\caption{Results comparison between the image sequence RCNN classifier (top row) and light curve RF classifier (bottom row). The left plot shows the accuracy on the simulated dataset as a function of the object magnitude and the number of images. The right plot shows the recall for each class available on the real dataset as a function of the number of samples presented to the classifier.}
\label{fig:model_comparison}
\end{figure*}

After completing the training process, we evaluated the image sequence classifier and the light curve classifier models both on the simulated dataset and on the real image dataset. The RCNN is evaluated both with and without fine-tuning. The light curve RF classifier performance is evaluated both using the simulated training set and the augmented training set. In the case of supernovae and asteroids, we define the first detection as the first point where the number of counts on the estimated light curve is five times higher than the counts error, then we build the input to the models as described in section \ref{inputs_to_model}. For the rest of the classes, the sequence starts at the first exposure. Figure \ref{fig:sim_image_performance} shows a comparison between the image sequence classifier (RCNN) and the light curve RF classifier, in terms of the accuracy evolution for simulated data as a function of the number of samples, images in case of the RCNN and light curve points in case of the light curve RF classifier, standard deviations were estimated by training 5 different models. Figure \ref{fig:real_image_performance} shows the average recall over all classes when the models are applied to real images, weighting every class equally. The curves in Figure \ref{fig:real_image_performance} show the average recall for real data using the models trained over the simulated dataset, as well as the RCNN with fine tuning and the RF classifier trained with the augmented training set, where the evaluation set does not include samples used to fine tune the model. The errors in this Figure were computed by taking 5 trials of 10 samples per class randomly selected to fine tune the model. Figure \ref{fig:model_comparison} shows the accuracy as a function of the magnitude of the simulated objects for both models, and the recall for real objects after fine tuning separated for each class, with standard deviations estimated using 5 different fine tuned models.

Figures \ref{fig:cm_sim_images} and \ref{fig:cm_sim_lightcurves} show confusion matrices for the RCNN and RF classifiers, respectively, on the simulated dataset after using all the points on each sequence to classify. Figures \ref{fig:cm_real_images} and \ref{fig:cm_real_lightcurves} show confusion matrices on the real dataset for both models when trained on simulated data only. Figures \ref{fig:cm_fine_tuning_images} and \ref{fig:cm_tuned_lightcurves} show confusion matrices for both models, after fine tuning the RCNN model and training the light curve RF classifier with the augmented training set.

\section{Discussion}
\label{sec:discussion}

\begin{figure}[!b]
\centerline{\includegraphics[width=0.5\textwidth]{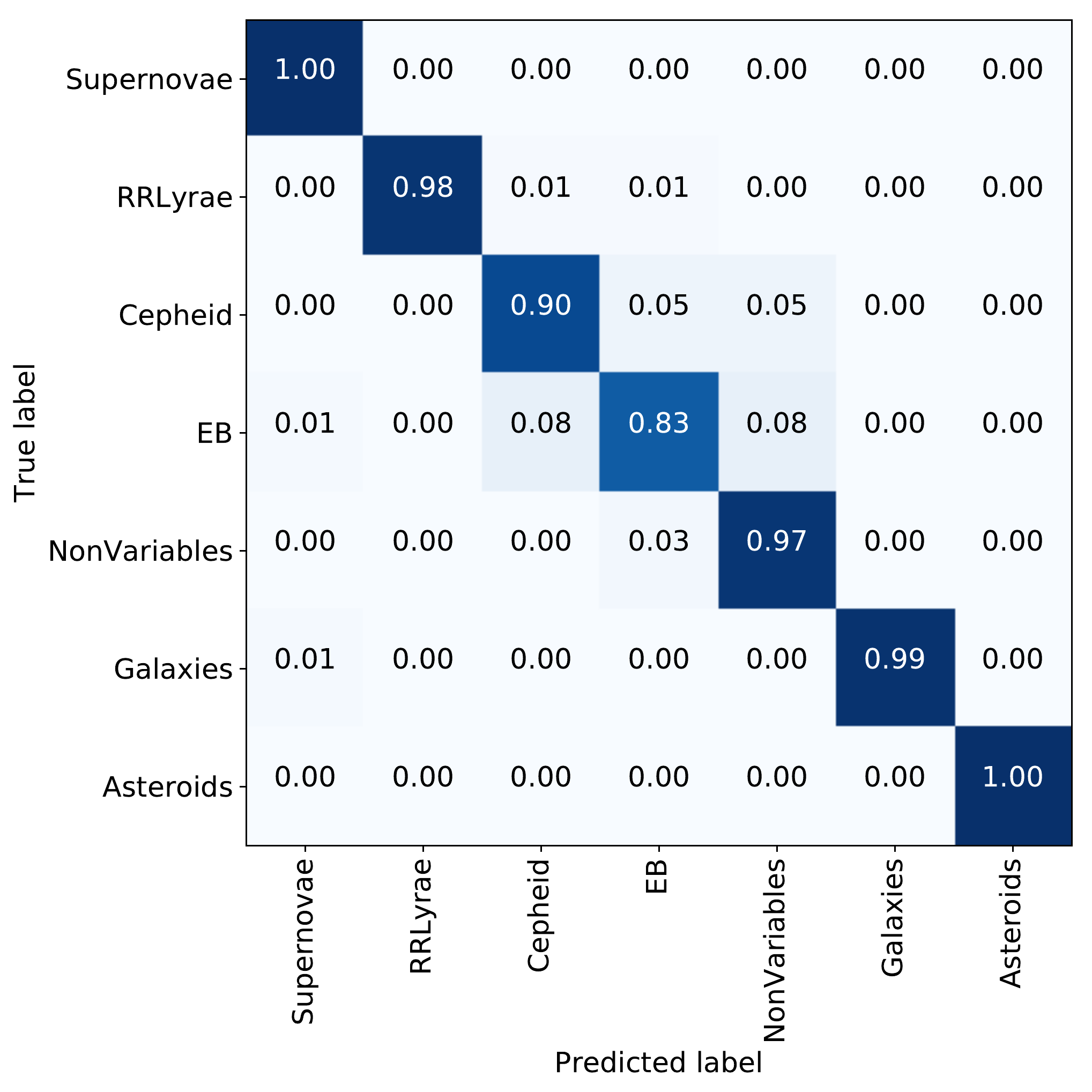}}
\caption{Confusion matrix on simulated test set obtained with the image sequence RCNN classifier using all samples available on each sequence to feed the neural network.}
\label{fig:cm_sim_images}
\end{figure}

\begin{figure}[!b]
\centerline{\includegraphics[width=0.5\textwidth]{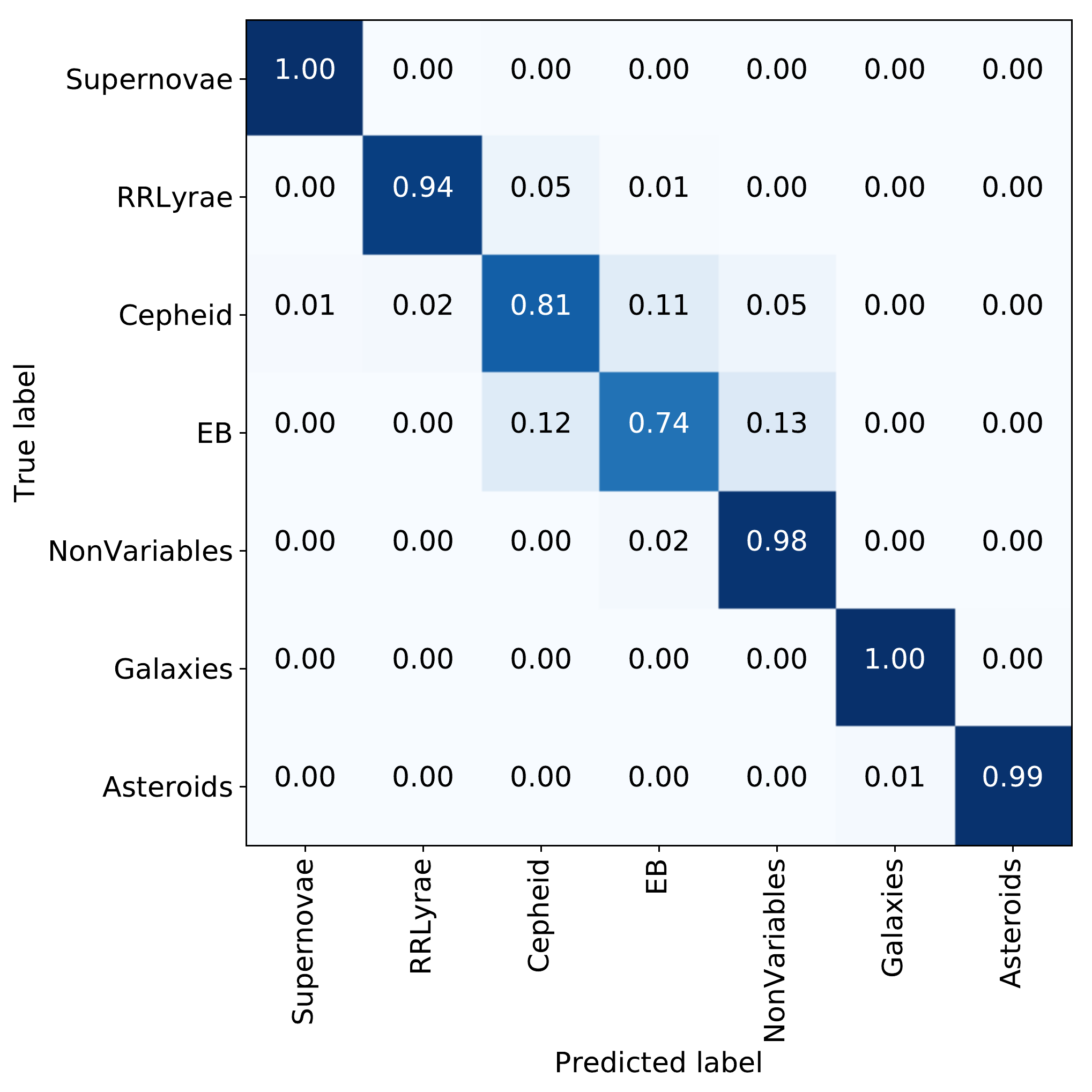}}
\caption{ Confusion matrix on simulated test set obtained with the light curve RF classifier  using all the points available on each light curve.}
\label{fig:cm_sim_lightcurves}
\end{figure}

\begin{figure}[!b]
\centerline{\includegraphics[width=0.5\textwidth]{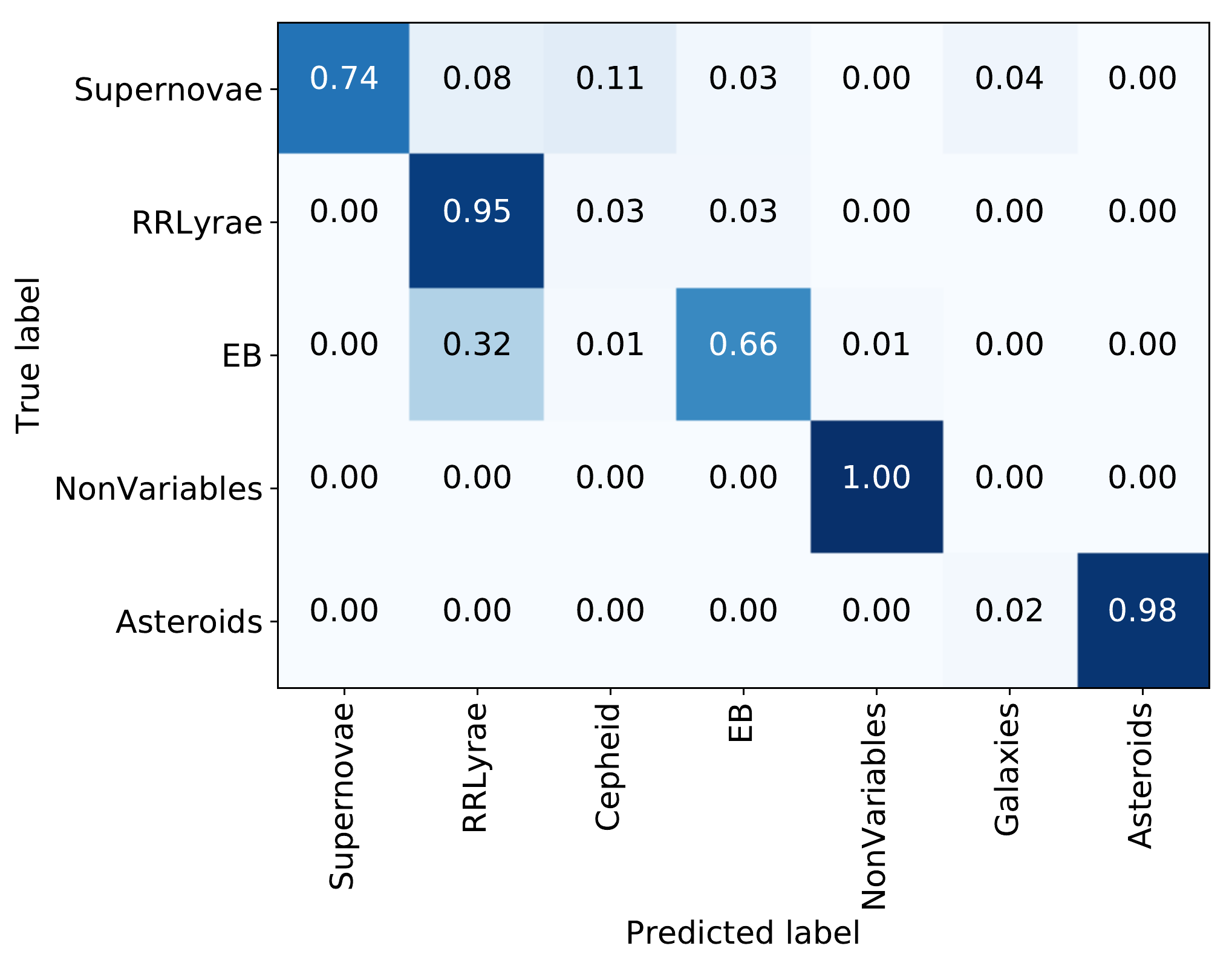}}
\caption{Confusion matrix on the HiTS real dataset obtained with the image sequence RCNN classifier without fine tuning.}
\label{fig:cm_real_images}
\end{figure}

\begin{figure}[!b]
\centerline{\includegraphics[width=0.5\textwidth]{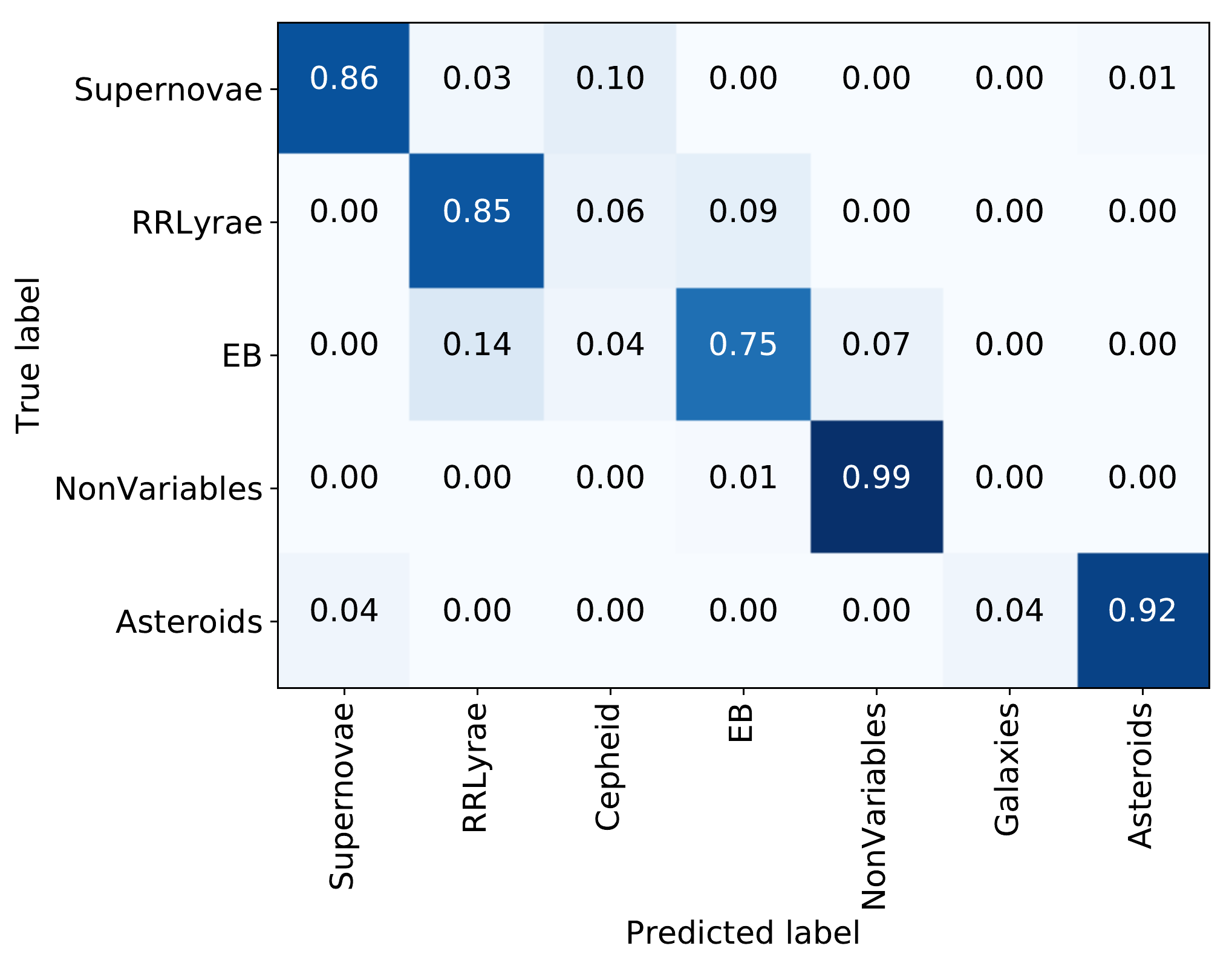}}
\caption{Confusion matrix on the HiTS real dataset obtained with the light curve RF classifier trained on simulated data.}
\label{fig:cm_real_lightcurves}
\end{figure}

\begin{figure}[!b]
\centerline{\includegraphics[width=0.5\textwidth]{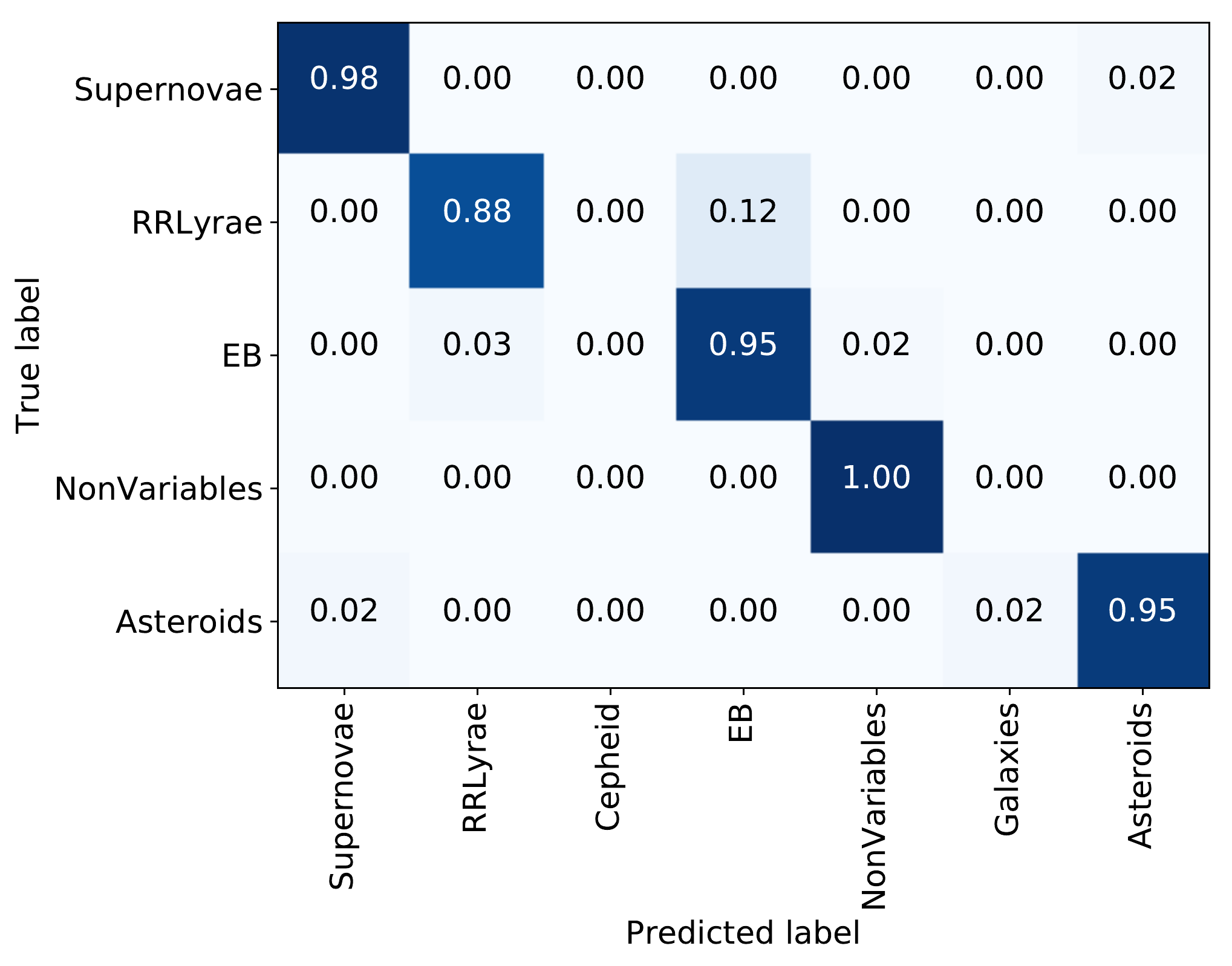}}
\caption{Confusion matrix on the HiTS real dataset obtained with the image sequence RCNN classifier with fine tuning. Real samples used to fine tune the model are not in this evaluation set.}
\label{fig:cm_fine_tuning_images}
\end{figure}

\begin{figure}[!b]
\centerline{\includegraphics[width=0.5\textwidth]{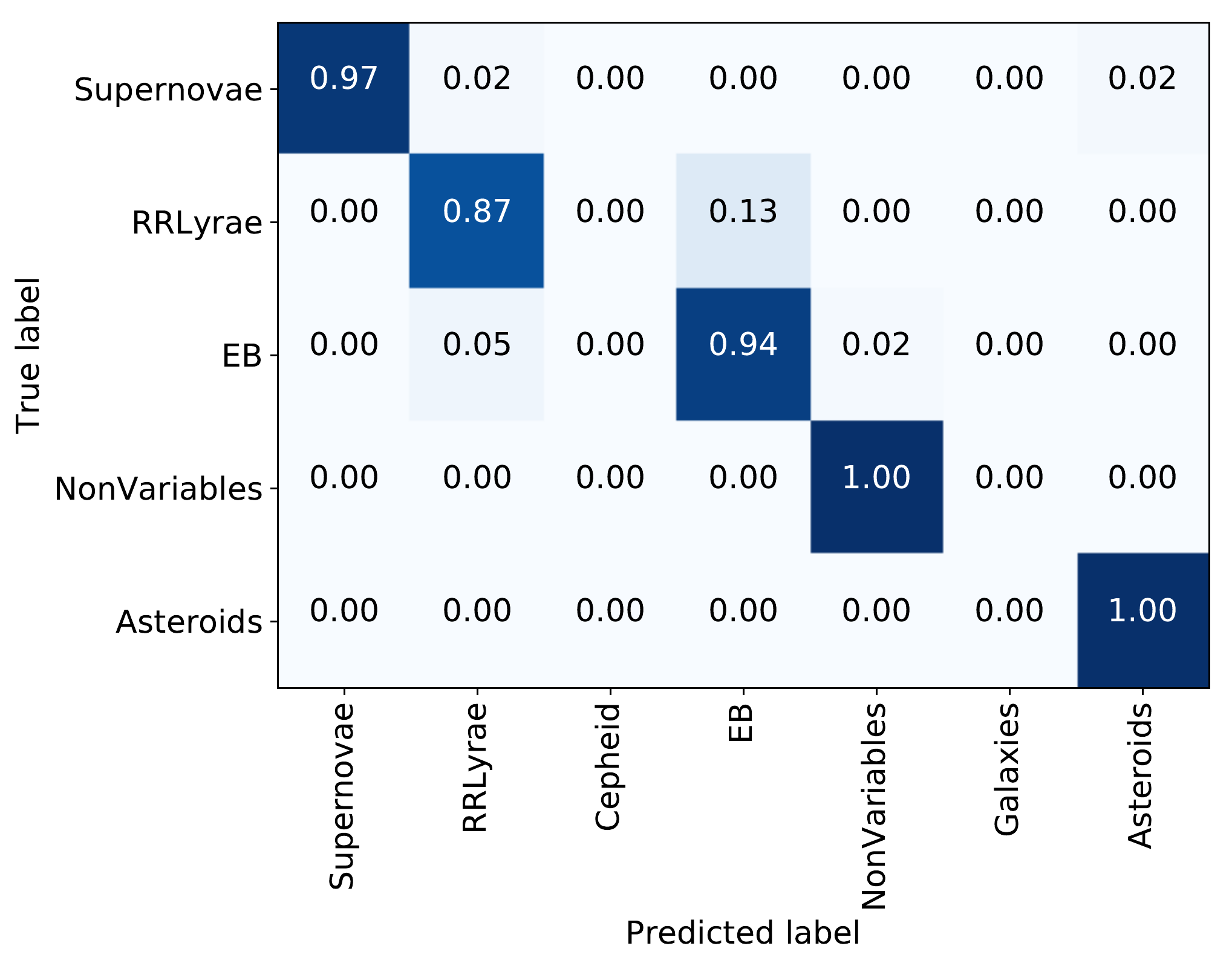}}
\caption{Confusion matrix on the HiTS real dataset obtained with the light curve RF classifier trained with the augmented training set. Real samples used to augment the training set are not in this evaluation set.}
\label{fig:cm_tuned_lightcurves}
\end{figure}

As can be observed in Figure \ref{fig:sim_image_performance}, the RCNN achieved a high accuracy on the simulated dataset as compared to the light curve RF classifier. Our model is particularly better on variable star classification (Cepheids, RR Lyrae and eclipsing binaries) as can be observed in Figures \ref{fig:cm_sim_images} and \ref{fig:cm_sim_lightcurves}. This result suggests that the image sequence classifier is able to retrieve the necessary information directly from images to solve the classification task. Since there is not much bias on each class apart from the light curve shape, the model must learn to perform some form of photometry and extract the flux generated by the object on the image. On the other hand, Figures \ref{fig:real_image_performance}, and confusion matrices in Figures \ref{fig:cm_real_images} and \ref{fig:cm_real_lightcurves} show that the light curve RF classifier outperforms the RCNN classifier on the real dataset without fine tuning. This might be explained by a kind of ``Domain Overfitting''. If the RCNN were more overfitted to the domain of the simulated images than the light curve RF classifier, then its generalization capacity to the domain of real images would be diminished. In fact, the light curve RF classifier achieved a lower accuracy on the simulated dataset, likely producing less overfitting to it, but obtained a higher accuracy on the real dataset.

The domain overfitting can be fixed by adjusting the models using a few real labeled data. Figure \ref{fig:real_image_performance}, and confusion matrices in Figures \ref{fig:cm_fine_tuning_images} and \ref{fig:cm_tuned_lightcurves} show that both models perform substantially better when a few real samples are used, achieving results comparable to those obtained with the simulated dataset. Notice in Figure \ref{fig:real_image_performance} the improvement in accuracy at low number of samples, which is the moment previous to the first detection in the cases of asteroids and supernovae. From this we can infer that the RCNN model is able to use additional information from the stamps than just the flux of the source, such as the presence of a host galaxy improving the initial guess between asteroid and supernovae for moments before the first detection. This effect can be seen on the right plot of Figure \ref{fig:model_comparison} where asteroids are highly miss-classified by the light curve RF classifier until their first detection (sample number 6). In appendix B, we illustrate some examples of the image sequence classifier applied to real supernovae, where we can clearly observe this effect once again. This effect seems to be more important for fainter sources, which tend to be more distant and have smaller angular size galaxies, as can be observed on the left plot of Figure \ref{fig:model_comparison}, where there is a clear improvement in accuracy for fainter sources when using the image sequence RCNN classifier in comparison to the light curve RF classifier.

\begin{figure*}[!t]
\centerline{\includegraphics[width=1.0\textwidth]{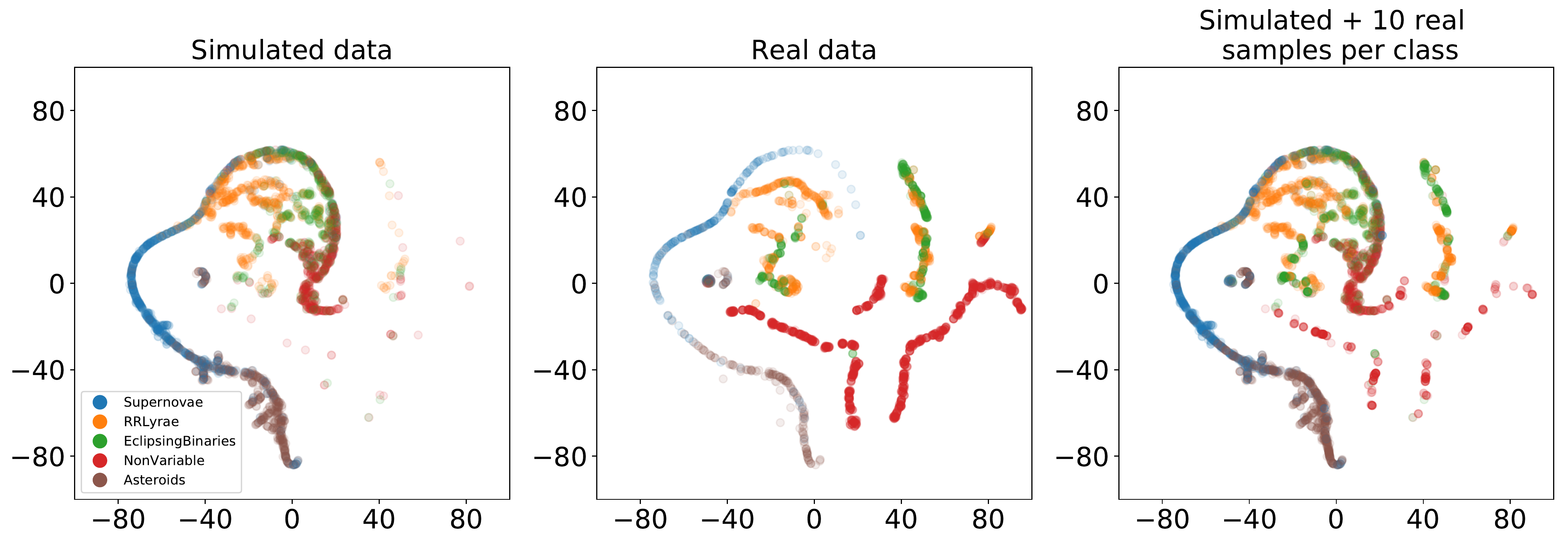}}
\caption{2D projection of simulated and real light curves, represented by the 6 most important FATS features, according to random forests, using t-SNE. The six features are \textit{MeanVariance, PercentDifferenceFluxPercentile, CAR\_mean, PercentAmplitude, Mean, Max\_slope}. 2000 light curves from the simulated dataset were used along with all the real light curves available, using a perplexity value of 40 and 1000 iterations.}
\label{fig:tsne}
\end{figure*}

It is worth to mention the fact that the simulations are good enough to classify correctly most of the real objects in the HiTS dataset. We can see in Figure \ref{fig:cm_real_images} that real non--variable objects are perfectly classified using the image sequence classifier, as well as the majority of supernovae (74\%), RR Lyrae (95\%), eclipsing binaries (66\%) and asteroids (98\%), resulting in an average recall of 85\% on the HiTS dataset. This means that the process of obtaining the right distribution that represents the physical effects of the light passing through the atmosphere, lenses and being captured by the CCD camera, by using the estimated exposure conditions, camera parameters and point spread function from empirical data is close enough to real conditions, making the simulation approach suitable to train a good model. The resulting model is used as a starting point to perform fine tuning with a few real samples, and get excellent results as shown in the confusion matrices in Figures \ref{fig:cm_fine_tuning_images} and \ref{fig:cm_tuned_lightcurves}.

Figure \ref{fig:tsne} shows a  2D projection of both the simulated and real light curves, using a technique called t-Distributed Stochastic Neighbor Embedding \cite[t-SNE][]{vanDerMaaten2008}. Each light curve is represented by a vector composed of the 6 most relevant features, according to random forests, and then projected to two-dimensions. The left and center plots show that there is partial overlap between the simulated and real light curves, particularly on supernovae, RR Lyrae and asteroids. On the other hand, there are stronger discrepancies for non--variables and eclipsing binaries, indicating some discordance between simulated and real datasets. The right plot in Figure \ref{fig:tsne} shows the 2D projection of the augmented dataset, i.e. the simulated dataset plus 10 real light curves per class randomly sampled. It can be clearly seen that this new distribution contains samples of eclipsing binaries and non--variables, among others, that are not present on the left plot. Each real sample added can be weighted with more importance through the fine-tuning phase of the RCNN by iterating these examples more times than the simulated ones, or making more copies of each real sample in the augmented training set in the case of the light curve RF classifier. 

For practical usage and deployment in an alert streaming classification scenario, our RCNN classifier presents some advantages compared to the light curve RF classifier. First, it is not necessary to compute neither the difference image nor the light curve, which avoids wrong subtractions due to alignment errors or PSF matching, it also reduces the amount of previous computation and information retrieval necessary to compute them. Second, for the RCNN the evaluation time increases linearly with the number of samples within the sequence, while for the light curve RF classifier the evaluation time depends on the complexity of the features used. Furthermore in the case of a streaming, with the proposed approach it is not necessary to recompute any feature when a new sample from a source is received, only the current sample and the $n_{w} - 1$ previous images, plus the network state is needed to evaluate the model. Third, the RCNN model is able to recognize other sources (stars or galaxies) from the image stamps beyond the flux from the main source. The RCNN model learns the distribution of these extra sources from the training set and applies it automatically when the evaluation is performed, with no need of additional image pre-processing, to create a flag or feature associated with these extra sources. This is particularly useful when a detailed catalog of nearby objects such as other stars or galaxies is not available. Fourth, our RCNN model trained using simulated images can be fine tuned when receiving real images from a stream, using a few labeled real images. On the other hand, the feature based light curve RF classifier still presents some desirable properties such as the interpretability of the features, the feature relevance for the classification task, easy implementation (except for the need to recompute features using the entire light curve at every time step) and so far a better generalization capacity without fine tuning. In a realistic setup, combining both approaches may improve the overall performance.

\section{Conclusions} 
\label{sec:conclusions}
We have proposed a new sequential classification model for astronomical objects based on a convolutional recurrent neural network, which uses directly sequences of images, and it does not require computing the light curve nor the difference image. Using empirical and astrophysical models of different astronomical objects, we simulated data from the High cadence Transient Survey (HiTS) to generate a synthetic dataset of images for seven classes of astronomical objects, considering realistic atmospheric conditions and camera specifications. This synthetic dataset was used to train the image sequence classifier (RCNN), then we evaluated the model on real images from the HiTS survey. We also performed fine tuning to adjust the models obtained with the simulated dataset using a few real samples per class. The results show that the proposed RCNN model is able to classify correctly the simulated test set, as well as the real dataset after performing fine tuning. To the best of our knowledge, this is the first time that a sequential classifier based on recurrent convolutional neural networks using sequences of images as inputs, and without computing the light curve or the difference images, has been proposed in time-domain astronomy. 

In order to assess the advantage of using images directly instead of light curves, we computed optimal photometry on the simulated images and trained a feature based light curve RF classifier using FATS features. Although the proposed RCNN model obtained similar results to those of the more traditional approach, it presents several advantages in an alert streaming classification scenario, such as no need for expensive pre-processing such as difference imaging and light curve computation, a faster evaluation, use of extra information aside from the source presented in the stamp such as the presence of a nearby galaxy, and an easy way to fine tune the model after receiving new images from the data streaming.

This work also shows that having images at the location of a transient event from before their first detection, can be very useful for the classification of astronomical alert streams. For example, in our proposed RCNN model having images of a supernova very early rise, before crossing a given threshold on the light curve, helps the classifier report better class probabilities.

Using realistic image simulations makes the domain adaptation problem from synthetic to real images easier to solve. As mentioned before, we were able to substantially improve the performance of our model by doing fine tuning with only 10 real samples per class. Using synthetic data is a reliable way to train models before acquiring real images from telescopes, as long as we have good light curve models available and the correct parameter distributions to represent well objects for a specific classification task and science goal. The fact that our RCNN classifier works well on real-world images after being trained with synthetic data and improves its performance after fine tuning, encourages us to use this methodology to train classifiers for new telescopes such as the Large Synoptic Survey Telescope \cite[LSST,][]{LSST1} and the Zwicky Transient Facility \cite[ZTF,][]{ZTF1}. This way, we could have a sequential classifier model available even before receiving real data, and as new data becomes available we can easily adjust our model.

\subsection{Future Work}
\label{future_work}

The results presented in this work could be improved by searching for the best hyperparameters of the recurrent convolutional neural network, as well as testing different architectures and optimization algorithms. Our current model only uses the image sequence preprocessed by zero points and sky brightness, and the dates of observation. Other observation conditions could be added to the model such as using the point spread function as an extra input image, or prior knowledge about the position of an object in the sky. For example, RR Lyrae and Cepheids are more likely to be found in the Milky Way plane than supernovae. Furthermore, the RCNN model could be combined somehow with the light curve RF classifier looking for an improvement in overall accuracy, along with other features related to the objects such as redshift or the presence of known nearby objects (e.g. galaxies or bright stars).

Since we want to be able to simulate other surveys and telescopes, we need to gather the relevant information about the observation conditions and camera specifications. Some telescopes such as the LSST are not operative yet and it is necessary to predict the observation conditions in order to simulate data. For this reason, a model invariant to observation conditions should be explored to increase robustness to prediction errors. We will explore simulating the sequence of images by randomizing observation conditions such as seeing, zero points, or airmass sampled from realistic distributions. Furthermore, our model is currently trained in band $g$, but it could be adapted to classify the image sequence combining information from more than one band, i.e., a multi-band image sequence classifier.

As future work, we propose to add more astronomical object classes, which means gathering more light curve models and adding them to the simulations, and use better priors for the presence of host galaxies for every object. We can also consider other effects associated with the CCD camera, such as hot pixels, bad columns and saturation effects, incorporating the real/bogus separation naturally into the proposed framework. 

\begin{acknowledgements}
Pablo Estevez, Francisco F\"orster, Pablo Huijse and Guillermo Cabrera-Vives acknowledge support from FONDECYT through grants
1171678, 11130228, 1170305 and 3160747, respectively. Ignacio Reyes gratefully acknowledges financial support from
CONICYT-PCHA through its national M.Sc. scholarship 2016
number 22162464. Rodrigo Carrasco Davis gratefully acknowledges financial support from
CONICYT-PCHA through its national M.Sc. scholarship 2018
number 22180964. The authors acknowledge support from the Chilean Ministry of Economy, Development, and Tourism’s Millennium Science Initiative through grant IC12009, awarded to the Millennium Institute of Astrophysics, MAS.
Francisco Förster acknowledge support from Center for Mathematical Modeling through Basal Project PFB-03.
Powered@NLHPC: this research  was partially supported by the supercomputing infrastructure of the NLHPC (ECM--02).
We thank the Nvidia GPU Research Center from University of Chile. 
\end{acknowledgements}



\bibliographystyle{pasa-mnras}
\bibliography{image_classification.bib}

\begin{thebibliography}{}
\makeatletter
\relax
\def\mn@urlcharsother{\let\do\@makeother \do\$\do\&\do\#\do\^\do\_\do\%\do\~}
\definecolor{darkblue}{rgb}{0,0,0.597656}
\def\mndoi{\begingroup\mn@urlcharsother \@ifnextchar [ {\mndoi@} {\mndoi@[]}}
\def\mndoi@[#1]#2{\def\@tempa{#1}\ifx\@tempa\@empty \href
  {http://dx.doi.org/#2} {\textcolor{darkblue}{doi:#2}}\else \href
  {http://dx.doi.org/#2} {\textcolor{darkblue}{#1}}\fi \endgroup}
\def\mn@eprint#1#2{\mn@eprint@#1:#2::\@nil}
\def\mn@eprint@arXiv#1{\href {http://arxiv.org/abs/#1} {{\tt arXiv:#1}}}
\def\mn@eprint@dblp#1{\href {http://dblp.uni-trier.de/rec/bibtex/#1.xml}
  {dblp:#1}}
\def\mn@eprint@#1:#2:#3:#4\@nil{\def\@tempa {#1}\def\@tempb {#2}\def\@tempc
  {#3}\ifx \@tempc \@empty \let \@tempc \@tempb \let \@tempb \@tempa \fi \ifx
  \@tempb \@empty \def\@tempb {arXiv}\fi \@ifundefined
  {mn@eprint@\@tempb}{\@tempb:\@tempc}{\expandafter \expandafter \csname
  mn@eprint@\@tempb\endcsname \expandafter{\@tempc}}}

\bibitem[\protect\citeauthoryear{Abdel-Hamid, r. Mohamed, Jiang, Deng, Penn  \&
  Yu}{Abdel-Hamid et~al.}{2014}]{audio2}
Abdel-Hamid O.,  r. Mohamed A.,  Jiang H.,  Deng L.,  Penn G.,   Yu D.,  2014,
  \mndoi [IEEE/ACM Transactions on Audio, Speech, and Language Processing]
  {10.1109/TASLP.2014.2339736}, 22, 1533

\bibitem[\protect\citeauthoryear{Belokurov, Evans  \& Du}{Belokurov
  et~al.}{2003}]{massive_lc_1}
Belokurov V.,  Evans N.~W.,   Du Y.~L.,  2003, \mndoi [Monthly Notices of the
  Royal Astronomical Society] {10.1046/j.1365-8711.2003.06512.x}, 341, 1373

\bibitem[\protect\citeauthoryear{Belokurov, Evans  \& Du}{Belokurov
  et~al.}{2004}]{massive_lc_2}
Belokurov V.,  Evans N.~W.,   Du Y.~L.,  2004, \mndoi [Monthly Notices of the
  Royal Astronomical Society] {10.1111/j.1365-2966.2004.07917.x}, 352, 233

\bibitem[\protect\citeauthoryear{Benavente, Protopapas  \& Pichara}{Benavente
  et~al.}{2017}]{Protopapas2_probabilistic}
Benavente P.,  Protopapas P.,   Pichara K.,  2017, The Astrophysical Journal,
  845, 147

\bibitem[\protect\citeauthoryear{Blanton et~al.,}{Blanton et~al.}{2017}]{SDSS}
Blanton M.~R.,  et~al., 2017, The Astronomical Journal, 154, 28

\bibitem[\protect\citeauthoryear{Bloom et~al.,}{Bloom
  et~al.}{2012}]{automatic_disc}
Bloom J.~S.,  et~al., 2012, Publications of the Astronomical Society of the
  Pacific, 124, 1175

\bibitem[\protect\citeauthoryear{Breiman}{Breiman}{2001}]{Breiman2001}
Breiman L.,  2001, \mndoi [Machine Learning] {10.1023/A:1010933404324}, 45, 5

\bibitem[\protect\citeauthoryear{Brink, Richards, Poznanski, Bloom, Rice,
  Negahban  \& Wainwright}{Brink et~al.}{2013}]{using_machine}
Brink H.,  Richards J.~W.,  Poznanski D.,  Bloom J.~S.,  Rice J.,  Negahban S.,
    Wainwright M.,  2013, \mndoi [Monthly Notices of the Royal Astronomical
  Society] {10.1093/mnras/stt1306}, 435, 1047

\bibitem[\protect\citeauthoryear{Cabrera-Vives, Reyes, Förster, Estévez  \&
  Maureira}{Cabrera-Vives et~al.}{2016}]{IJCNN1}
Cabrera-Vives G.,  Reyes I.,  Förster F.,  Estévez P.~A.,   Maureira J.~C.,
  2016, in 2016 International Joint Conference on Neural Networks (IJCNN). pp
  251--258, \mndoi{10.1109/IJCNN.2016.7727206}

\bibitem[\protect\citeauthoryear{Cabrera-Vives, Reyes, Förster, Estévez  \&
  Maureira}{Cabrera-Vives et~al.}{2017}]{deep_hits}
Cabrera-Vives G.,  Reyes I.,  Förster F.,  Estévez P.~A.,   Maureira J.-C.,
  2017, The Astrophysical Journal, 836, 97

\bibitem[\protect\citeauthoryear{Castro, Protopapas  \& Pichara}{Castro
  et~al.}{2018}]{Protopapas1_features}
Castro N.,  Protopapas P.,   Pichara K.,  2018, The Astronomical Journal, 155,
  16

\bibitem[\protect\citeauthoryear{{Chambers} et~al.,}{{Chambers}
  et~al.}{2016}]{2016arXiv161205560C}
{Chambers} K.~C.,  et~al., 2016, preprint, \href
  {http://adsabs.harvard.edu/abs/2016arXiv161205560C} {} (\mn@eprint {arXiv}
  {1612.05560})

\bibitem[\protect\citeauthoryear{Charnock \& Moss}{Charnock \&
  Moss}{2017}]{lc_charnock}
Charnock T.,  Moss A.,  2017, The Astrophysical Journal Letters, 837, L28

\bibitem[\protect\citeauthoryear{Diehl}{Diehl}{2012}]{DECAM}
Diehl T.,  2012, 37, 1332

\bibitem[\protect\citeauthoryear{Donahue, Hendricks, Rohrbach, Venugopalan,
  Guadarrama, Saenko  \& Darrell}{Donahue et~al.}{2017}]{video1}
Donahue J.,  Hendricks L.~A.,  Rohrbach M.,  Venugopalan S.,  Guadarrama S.,
  Saenko K.,   Darrell T.,  2017, \mndoi [IEEE Transactions on Pattern Analysis
  and Machine Intelligence] {10.1109/TPAMI.2016.2599174}, 39, 677

\bibitem[\protect\citeauthoryear{Feast, Menzies, Matsugana  \& Whitelock}{Feast
  et~al.}{2014}]{milky2}
Feast M.~W.,  Menzies J.~W.,  Matsugana N.,   Whitelock P.~A.,  2014, Nature,
  509, 342

\bibitem[\protect\citeauthoryear{{F{\"o}rster} et~al.,}{{F{\"o}rster}
  et~al.}{2018}]{2018NatAs.tmp..122F}
{F{\"o}rster} F.,  et~al., 2018, \mndoi [Nature Astronomy]
  {10.1038/s41550-018-0563-4}, \href
  {http://adsabs.harvard.edu/abs/2018NatAs.tmp..122F} {}

\bibitem[\protect\citeauthoryear{Fukushima}{Fukushima}{1980}]{Original_cnn}
Fukushima K.,  1980, \mndoi [Biological Cybernetics] {10.1007/BF00344251}, 36,
  193

\bibitem[\protect\citeauthoryear{Förster et~al.,}{Förster
  et~al.}{2016}]{HiTS}
Förster F.,  et~al., 2016, The Astrophysical Journal, 832, 155

\bibitem[\protect\citeauthoryear{George \& Huerta}{George \&
  Huerta}{2017}]{gravitational_waves}
George D.,  Huerta E.~A.,  2017, \mndoi [Physics Letters B]
  {10.1016/j.physletb.2017.12.053}, 778, 64

\bibitem[\protect\citeauthoryear{Gers, Schmidhuber  \& Cummins}{Gers
  et~al.}{1999}]{lstm2}
Gers F.~A.,  Schmidhuber J.,   Cummins F.,  1999, Neural Computation, 12, 2451

\bibitem[\protect\citeauthoryear{Goldberg \& Hirst}{Goldberg \&
  Hirst}{2017}]{nlpbook}
Goldberg Y.,  Hirst G.,  2017, Neural Network Methods in Natural Language
  Processing.
Morgan \& Claypool Publishers

\bibitem[\protect\citeauthoryear{Graves, r. Mohamed  \& Hinton}{Graves
  et~al.}{2013}]{lstm_speech}
Graves A.,  r. Mohamed A.,   Hinton G.,  2013, in 2013 IEEE International
  Conference on Acoustics, Speech and Signal Processing. pp 6645--6649,
  \mndoi{10.1109/ICASSP.2013.6638947}

\bibitem[\protect\citeauthoryear{Hartman, Bersier, Stanek, Beaulieu, Kaluzny,
  Marquette, Stetson  \& Czerny}{Hartman et~al.}{2006}]{Cepheids}
Hartman J.~D.,  Bersier D.,  Stanek K.~Z.,  Beaulieu J.~P.,  Kaluzny J.,
  Marquette J.,  Stetson P.~B.,   Czerny A.~S.,  2006, \mndoi [Monthly Notices
  of the Royal Astronomical Society] {10.1111/j.1365-2966.2006.10764.x}, 371,
  1405

\bibitem[\protect\citeauthoryear{Hochreiter \& Schmidhuber}{Hochreiter \&
  Schmidhuber}{1997}]{lstm1}
Hochreiter S.,  Schmidhuber J.,  1997, Neural computation, 9, 1735

\bibitem[\protect\citeauthoryear{{Hsiao}, {Conley}, {Howell}, {Sullivan},
  {Pritchet}, {Carlberg}, {Nugent}  \& {Phillips}}{{Hsiao}
  et~al.}{2007}]{2007ApJ...663.1187H}
{Hsiao} E.~Y.,  {Conley} A.,  {Howell} D.~A.,  {Sullivan} M.,  {Pritchet}
  C.~J.,  {Carlberg} R.~G.,  {Nugent} P.~E.,   {Phillips} M.~M.,  2007, \mndoi
  [\apj] {10.1086/518232}, \href
  {http://adsabs.harvard.edu/abs/2007ApJ...663.1187H} {663, 1187}

\bibitem[\protect\citeauthoryear{{Ioffe}}{{Ioffe}}{2017}]{2017arXiv170203275I}
{Ioffe} S.,  2017, preprint, \href
  {http://adsabs.harvard.edu/abs/2017arXiv170203275I} {} (\mn@eprint {arXiv}
  {1702.03275})

\bibitem[\protect\citeauthoryear{Ioffe \& Szegedy}{Ioffe \&
  Szegedy}{2015}]{BatchNorm}
Ioffe S.,  Szegedy C.,  2015, in Proceedings of the 32Nd International
  Conference on International Conference on Machine Learning - Volume 37.
  ICML'15.
JMLR.org, pp 448--456

\bibitem[\protect\citeauthoryear{Ivezic et~al.,}{Ivezic et~al.}{2008}]{LSST1}
Ivezic Z.,  et~al., 2008, ArXiv e-prints arXiv:0805.2366v4, \href
  {http://adsabs.harvard.edu/abs/2008arXiv0805.2366I} {}

\bibitem[\protect\citeauthoryear{{Kim}, {Protopapas}, {Bailer-Jones}, {Byun},
  {Chang}, {Marquette}  \& {Shin}}{{Kim} et~al.}{2014}]{2014A&A...566A..43K}
{Kim} D.-W.,  {Protopapas} P.,  {Bailer-Jones} C.~A.~L.,  {Byun} Y.-I.,
  {Chang} S.-W.,  {Marquette} J.-B.,   {Shin} M.-S.,  2014, \mndoi [\aap]
  {10.1051/0004-6361/201323252}, \href
  {http://adsabs.harvard.edu/abs/2014A%26A...566A..43K} {566, A43}

\bibitem[\protect\citeauthoryear{Kimura, Takahashi, Tanaka, Yasuda, Ueda  \&
  Yoshida}{Kimura et~al.}{2017}]{cnn_japon}
Kimura A.,  Takahashi I.,  Tanaka M.,  Yasuda N.,  Ueda N.,   Yoshida N.,
  2017, in 2017 IEEE 37th International Conference on Distributed Computing
  Systems Workshops (ICDCSW). pp 354--359, \mndoi{10.1109/ICDCSW.2017.47}

\bibitem[\protect\citeauthoryear{Krizhevsky, Sutskever  \& Hinton}{Krizhevsky
  et~al.}{2012}]{AlexNet}
Krizhevsky A.,  Sutskever I.,   Hinton G.~E.,  2012, in Pereira F.,  Burges C.
  J.~C.,  Bottou L.,   Weinberger K.~Q.,  eds, , Advances in Neural Information
  Processing Systems 25.
Curran Associates, Inc., pp 1097--1105

\bibitem[\protect\citeauthoryear{Lee, Pham, Largman  \& Ng}{Lee
  et~al.}{2009}]{audio1}
Lee H.,  Pham P.,  Largman Y.,   Ng A.~Y.,  2009, in Bengio Y.,  Schuurmans D.,
   Lafferty J.~D.,  Williams C. K.~I.,   Culotta A.,  eds, , Advances in Neural
  Information Processing Systems 22.
Curran Associates, Inc., pp 1096--1104

\bibitem[\protect\citeauthoryear{{Lipton}, {Berkowitz}  \& {Elkan}}{{Lipton}
  et~al.}{2015}]{Recurrent}
{Lipton} Z.~C.,  {Berkowitz} J.,   {Elkan} C.,  2015, ArXiv e-prints
  arXiv:1506.00019

\bibitem[\protect\citeauthoryear{{Lomb}}{{Lomb}}{1976}]{1976Ap&SS..39..447L}
{Lomb} N.~R.,  1976, \mndoi [\apss] {10.1007/BF00648343}, \href
  {http://adsabs.harvard.edu/abs/1976Ap%26SS..39..447L} {39, 447}

\bibitem[\protect\citeauthoryear{{Mahabal}, {Sheth}, {Gieseke}, {Pai},
  {Djorgovski}, {Drake}, {Graham}  \& {the CSS/CRTS/PTF
  Collaboration}}{{Mahabal} et~al.}{2017}]{learnt}
{Mahabal} A.,  {Sheth} K.,  {Gieseke} F.,  {Pai} A.,  {Djorgovski} S.~G.,
  {Drake} A.,  {Graham} M.,   {the CSS/CRTS/PTF Collaboration} 2017, in 2017
  IEEE Symposium Series on Computational Intelligence (SSCI). pp~1--8,
  \mndoi{10.1109/SSCI.2017.8280984}

\bibitem[\protect\citeauthoryear{{Mart{\'{\i}}nez-Palomera}
  et~al.,}{{Mart{\'{\i}}nez-Palomera} et~al.}{2018}]{2018arXiv180900763M}
{Mart{\'{\i}}nez-Palomera} J.,  et~al., 2018, preprint, \href
  {http://adsabs.harvard.edu/abs/2018arXiv180900763M} {} (\mn@eprint {arXiv}
  {1809.00763})

\bibitem[\protect\citeauthoryear{Medina et~al.,}{Medina
  et~al.}{2018}]{medina2018}
Medina G.~E.,  et~al., 2018, The Astrophysical Journal, 855, 43

\bibitem[\protect\citeauthoryear{Moriya et~al.,}{Moriya et~al.}{2017}]{Moriya}
Moriya T.~J.,  et~al., 2017, \mndoi [Monthly Notices of the Royal Astronomical
  Society] {10.1093/mnras/stw3225}, 466, 2085

\bibitem[\protect\citeauthoryear{Nair \& Hinton}{Nair \& Hinton}{2010}]{ReLU}
Nair V.,  Hinton G.~E.,  2010, in Proceedings of the 27th International
  Conference on International Conference on Machine Learning. ICML'10.
Omnipress, USA, pp 807--814

\bibitem[\protect\citeauthoryear{Narayan et~al.,}{Narayan
  et~al.}{2018}]{ANTARES}
Narayan G.,  et~al., 2018, The Astrophysical Journal Supplement Series, 236, 9

\bibitem[\protect\citeauthoryear{{Naul}, {Bloom}, {P{\'e}rez}  \& {van der
  Walt}}{{Naul} et~al.}{2018}]{2018NatAs...2..151N}
{Naul} B.,  {Bloom} J.~S.,  {P{\'e}rez} F.,   {van der Walt} S.,  2018, \mndoi
  [Nature Astronomy] {10.1038/s41550-017-0321-z}, \href
  {http://adsabs.harvard.edu/abs/2018NatAs...2..151N} {2, 151}

\bibitem[\protect\citeauthoryear{Naylor}{Naylor}{1998}]{Naylor}
Naylor T.,  1998, \mndoi [Monthly Notices of the Royal Astronomical Society]
  {10.1046/j.1365-8711.1998.01314.x}, 296, 339

\bibitem[\protect\citeauthoryear{Ngeow, Gieren  \& Klein}{Ngeow
  et~al.}{2013}]{distance1}
Ngeow C.,  Gieren W.,   Klein C.,  2013, \mndoi [Proceedings of the
  International Astronomical Union] {10.1017/S1743921313014208}, 9, 123–128

\bibitem[\protect\citeauthoryear{Nun, Protopapas, Sim, Zhu, Dave, Castro  \&
  Pichara}{Nun et~al.}{2015}]{FATS}
Nun I.,  Protopapas P.,  Sim B.,  Zhu M.,  Dave R.,  Castro N.,   Pichara K.,
  2015, ArXiv e-prints arXiv:1506.00010, \href
  {http://adsabs.harvard.edu/abs/2015arXiv150600010N} {}

\bibitem[\protect\citeauthoryear{Oquab, Bottou, Laptev  \& Sivic}{Oquab
  et~al.}{2014}]{6909618}
Oquab M.,  Bottou L.,  Laptev I.,   Sivic J.,  2014, in 2014 IEEE Conference on
  Computer Vision and Pattern Recognition. pp 1717--1724,
  \mndoi{10.1109/CVPR.2014.222}

\bibitem[\protect\citeauthoryear{Pichara \& Protopapas}{Pichara \&
  Protopapas}{2013}]{0004-637X-777-2-83}
Pichara K.,  Protopapas P.,  2013, The Astrophysical Journal, 777, 83

\bibitem[\protect\citeauthoryear{Protopapas}{Protopapas}{2017}]{protopapas4_lstm}
Protopapas P.,  2017, in American Astronomical Society Meeting Abstracts \#230.
  p. 104.03

\bibitem[\protect\citeauthoryear{Rasmussen \& Williams}{Rasmussen \&
  Williams}{2005}]{GaussianProcess}
Rasmussen C.~E.,  Williams C. K.~I.,  2005, Gaussian Processes for Machine
  Learning (Adaptive Computation and Machine Learning).
The MIT Press

\bibitem[\protect\citeauthoryear{Reddi, Kale  \& Kumar}{Reddi
  et~al.}{2018}]{j.2018on}
Reddi S.~J.,  Kale S.,   Kumar S.,  2018, in International Conference on
  Learning Representations. \url {https://openreview.net/forum?id=ryQu7f-RZ}

\bibitem[\protect\citeauthoryear{Riess et~al.,}{Riess
  et~al.}{1998}]{1998_supernovae}
Riess A.~G.,  et~al., 1998, The Astronomical Journal, 116, 1009

\bibitem[\protect\citeauthoryear{Sainath, Vinyals, Senior  \& Sak}{Sainath
  et~al.}{2015}]{video2}
Sainath T.~N.,  Vinyals O.,  Senior A.,   Sak H.,  2015, in 2015 IEEE
  International Conference on Acoustics, Speech and Signal Processing (ICASSP).
  pp 4580--4584, \mndoi{10.1109/ICASSP.2015.7178838}

\bibitem[\protect\citeauthoryear{Schmidt et~al.,}{Schmidt
  et~al.}{1998}]{supernovae_2}
Schmidt B.~P.,  et~al., 1998, The Astrophysical Journal, 507, 46

\bibitem[\protect\citeauthoryear{Sedaghat \& Mahabal}{Sedaghat \&
  Mahabal}{2017}]{difference}
Sedaghat N.,  Mahabal A.,  2017, ArXiv e-prints arXiv:1710.01422

\bibitem[\protect\citeauthoryear{Sesar et~al.,}{Sesar et~al.}{2010}]{RRLyrae}
Sesar B.,  et~al., 2010, The Astrophysical Journal, 708, 717

\bibitem[\protect\citeauthoryear{Shallue \& Vanderburg}{Shallue \&
  Vanderburg}{2018}]{exoplanets}
Shallue C.,  Vanderburg A.,  2018, The Astronomical Journal, 155, 94

\bibitem[\protect\citeauthoryear{Smith et~al.,}{Smith et~al.}{2014}]{ZTF1}
Smith R.~M.,  et~al., 2014, \mndoi [Proc. SPIE 9147, Ground-based and Airborne
  Instrumentation for Astronomy V] {10.1117/12.2070014}

\bibitem[\protect\citeauthoryear{Sutskever, Vinyals  \& Le}{Sutskever
  et~al.}{2014}]{translation1}
Sutskever I.,  Vinyals O.,   Le Q.~V.,  2014, in Ghahramani Z.,  Welling M.,
  Cortes C.,  Lawrence N.~D.,   Weinberger K.~Q.,  eds, , Advances in Neural
  Information Processing Systems 27.
Curran Associates, Inc., pp 3104--3112

\bibitem[\protect\citeauthoryear{Szegedy et~al.,}{Szegedy
  et~al.}{2015}]{GoogleNet}
Szegedy C.,  et~al., 2015, in 2015 IEEE Conference on Computer Vision and
  Pattern Recognition (CVPR). pp~1--9, \mndoi{10.1109/CVPR.2015.7298594}

\bibitem[\protect\citeauthoryear{{Yosinski}, {Clune}, {Bengio}  \&
  {Lipson}}{{Yosinski} et~al.}{2014}]{2014arXiv1411.1792Y}
{Yosinski} J.,  {Clune} J.,  {Bengio} Y.,   {Lipson} H.,  2014, in Proceedings
  of the 27th International Conference on Neural Information Processing
  Systems. MIT Press, pp 3320--3328

\bibitem[\protect\citeauthoryear{Yuan, He, Macri, Long  \& Huang}{Yuan
  et~al.}{2017}]{1538-3881-153-4-170}
Yuan W.,  He S.,  Macri L.~M.,  Long J.,   Huang J.~Z.,  2017, The Astronomical
  Journal, 153, 170

\bibitem[\protect\citeauthoryear{Zhao, Ali  \& van~der Smagt}{Zhao
  et~al.}{2017a}]{video3}
Zhao R.,  Ali H.,   van~der Smagt P.,  2017a, in 2017 IEEE/RSJ International
  Conference on Intelligent Robots and Systems (IROS). pp 4260--4267,
  \mndoi{10.1109/IROS.2017.8206288}

\bibitem[\protect\citeauthoryear{Zhao, Jin  \& Hu}{Zhao
  et~al.}{2017b}]{RCNN_speech}
Zhao Y.,  Jin X.,   Hu X.,  2017b, in 2017 IEEE International Conference on
  Acoustics, Speech and Signal Processing (ICASSP). pp 5300--5304,
  \mndoi{10.1109/ICASSP.2017.7953168}

\bibitem[\protect\citeauthoryear{van~der Maaten \& Hinton}{van~der Maaten \&
  Hinton}{2008}]{vanDerMaaten2008}
van~der Maaten L.,  Hinton G.,  2008, Journal of Machine Learning Research, 9,
  2579

\makeatother
\end{thebibliography}

\newpage
\onecolumn

\section*{Appendix A: Simulation examples and comparison with real images}
\label{appendix_A}

\begin{figure}[!h]
\centerline{\includegraphics[width=1.0\textwidth]{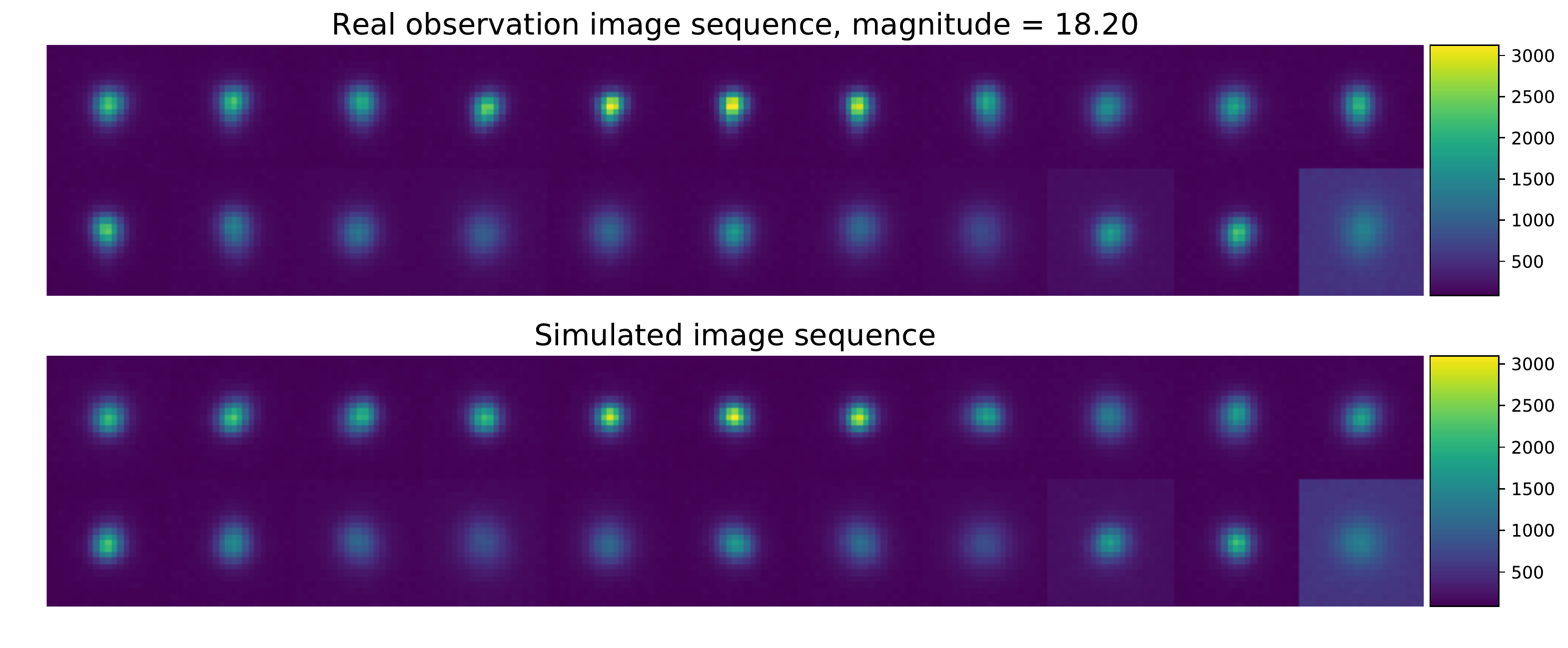}}
\caption{Image simulation example 1}
\label{fig:comp1}
\end{figure}

\begin{figure}[!h]
\centerline{\includegraphics[width=1.0\textwidth]{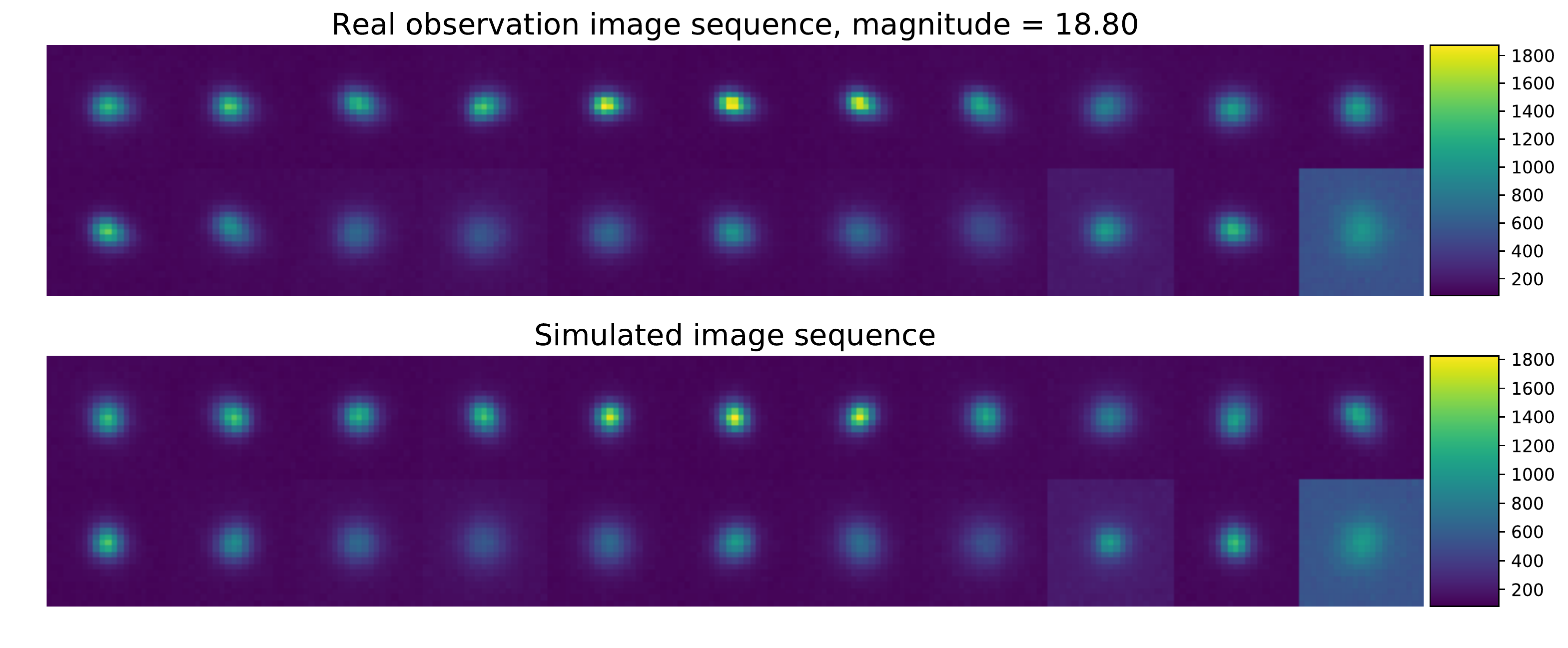}}
\caption{Image simulation example 2}
\label{fig:comp2}
\end{figure}

\begin{figure}[!t]
\centerline{\includegraphics[width=1.0\textwidth]{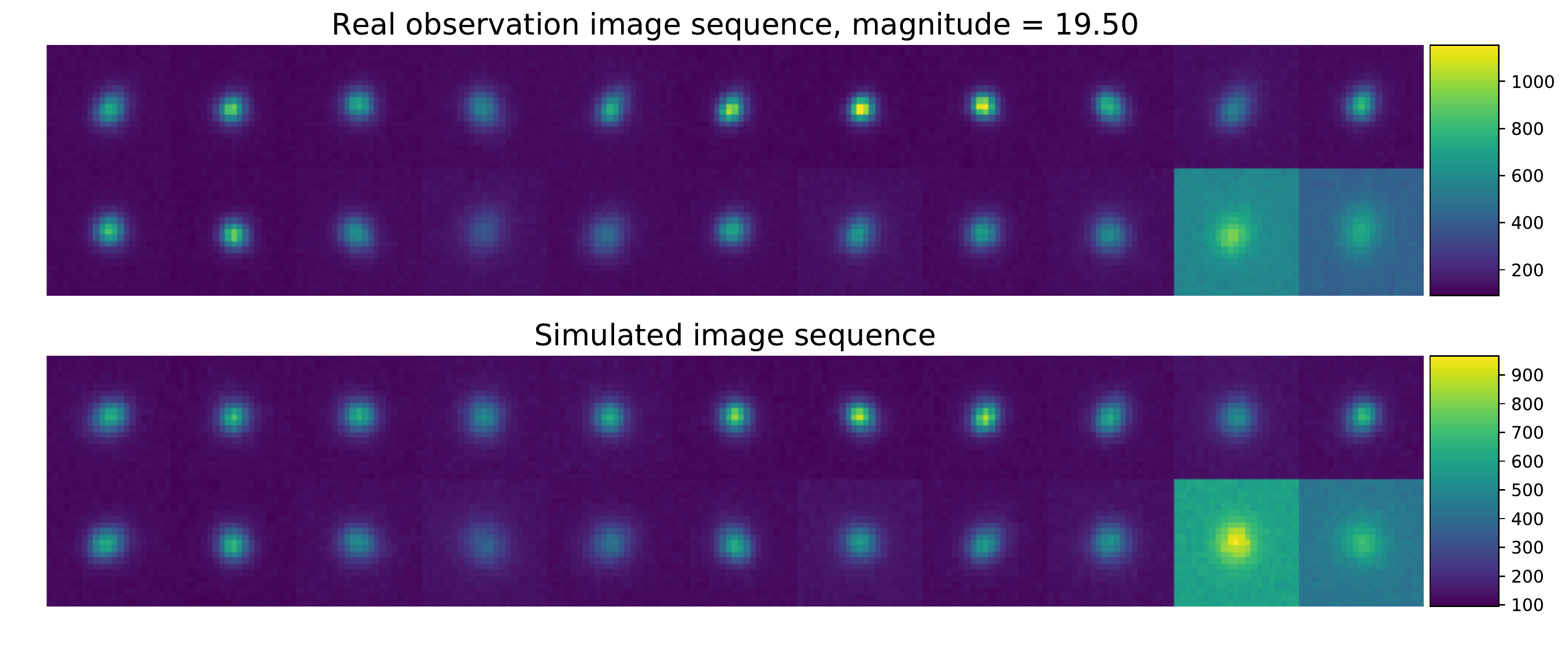}}
\caption{Image simulation example 3}
\label{fig:comp3}
\end{figure}

In Figs. \ref{fig:comp1} to \ref{fig:comp3}, we show examples of simulated images compared to the real ones. Given a real non--variable source with a known magnitude, we use the same dates where the source was observed and the observation conditions to simulate a non--variable source. Sample time goes from left to right, top to bottom. Since the estimated point spread function used for simulations is computed by averaging single PSFs, simulated images have bigger PSFs than real ones. As can be observed, the pixel distribution on the simulated images is very close to the real ones. 

\newpage
\section*{Appendix B: Examples of the image sequence classifier model working on real supernovae}
\label{appendix_B}

\begin{figure}[!h]
\centerline{\includegraphics[width=0.88\textwidth]{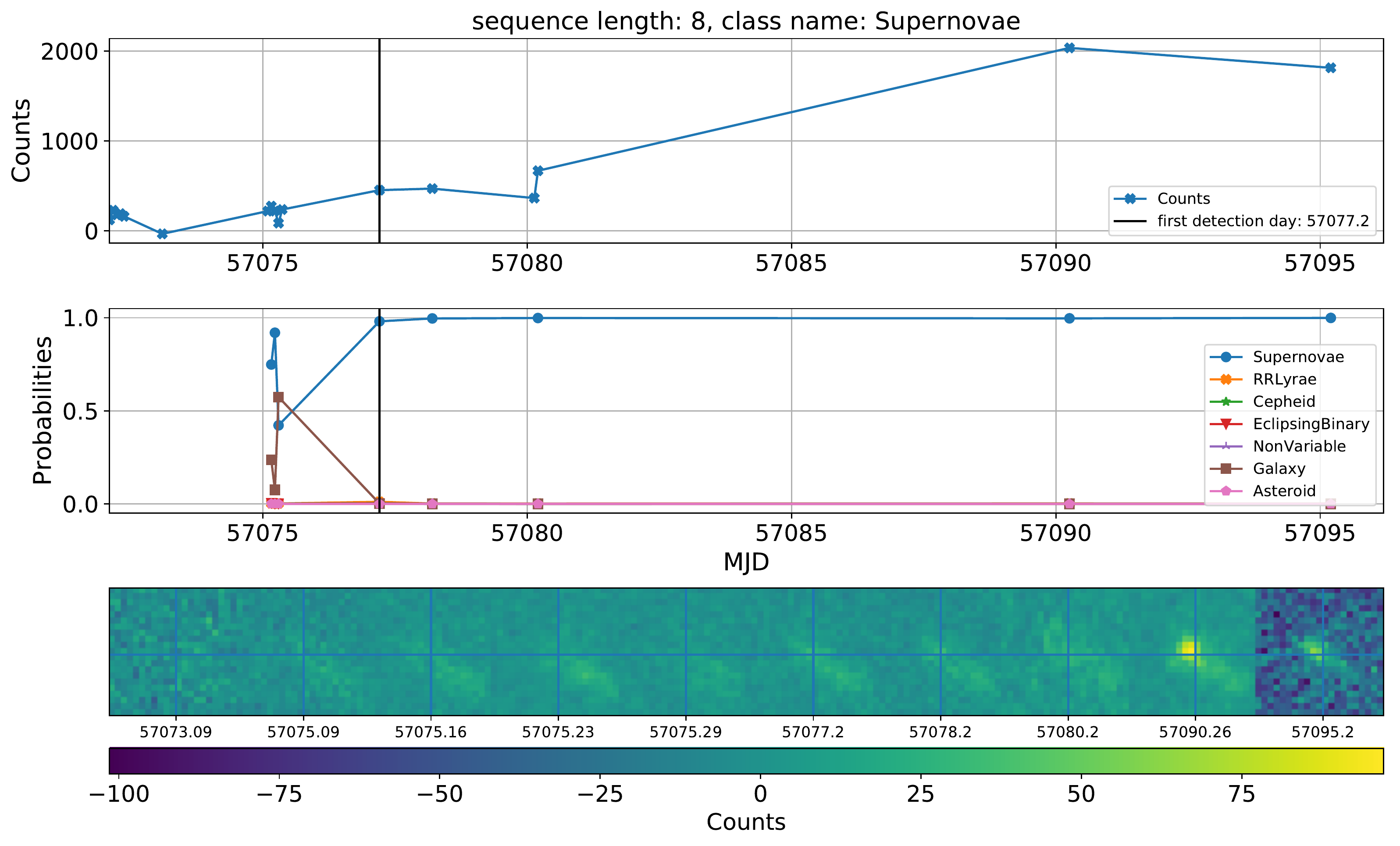}}
\caption{Example 1 of SN classification using the image sequence classifier}
\label{fig:ex1}
\end{figure}

\begin{figure}[!h]
\centerline{\includegraphics[width=0.88\textwidth]{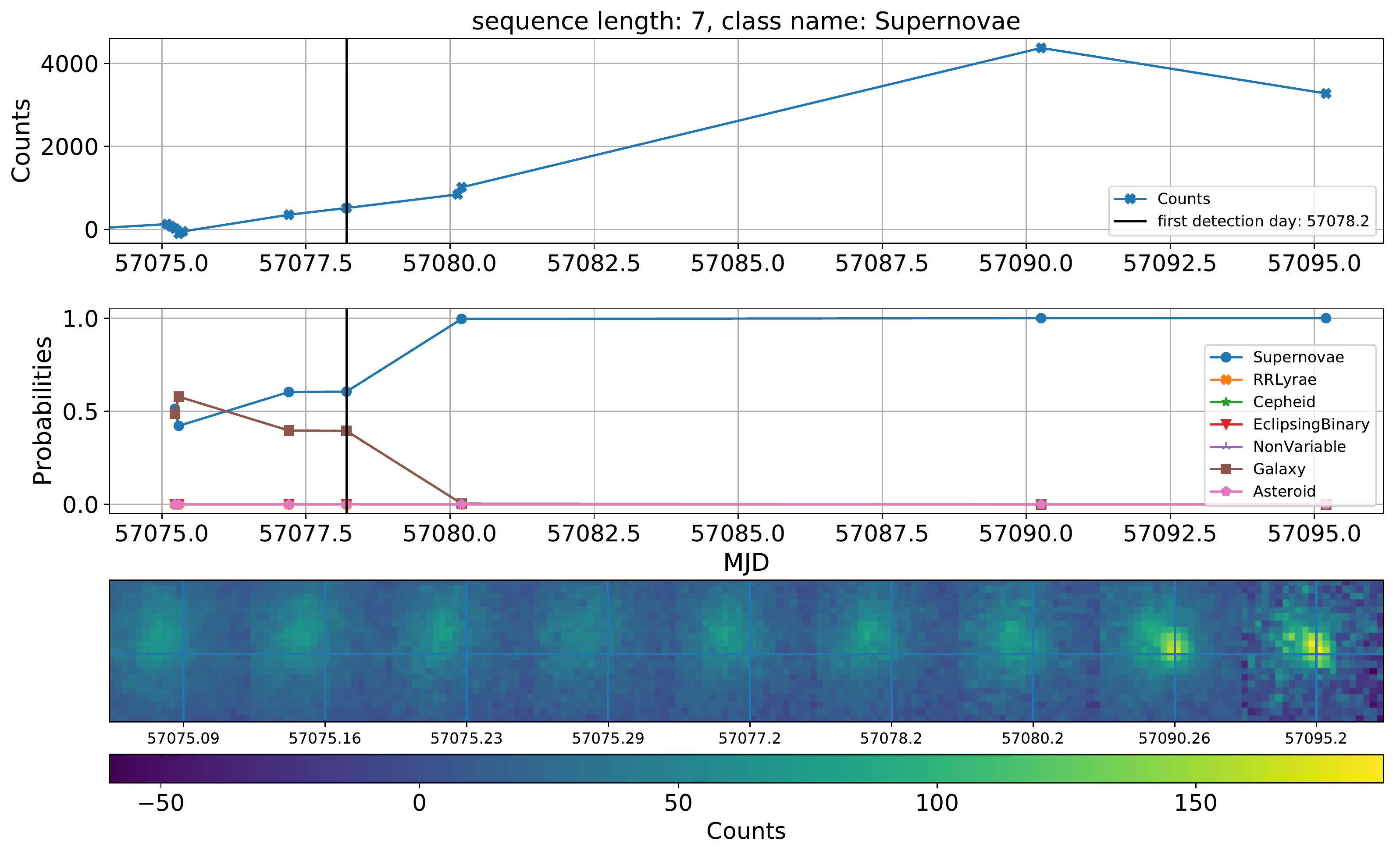}}
\caption{Example 2 of SN classification using the image sequence classifier}
\label{fig:ex2}
\end{figure}

\begin{figure}[!t]
\centerline{\includegraphics[width=0.88\textwidth]{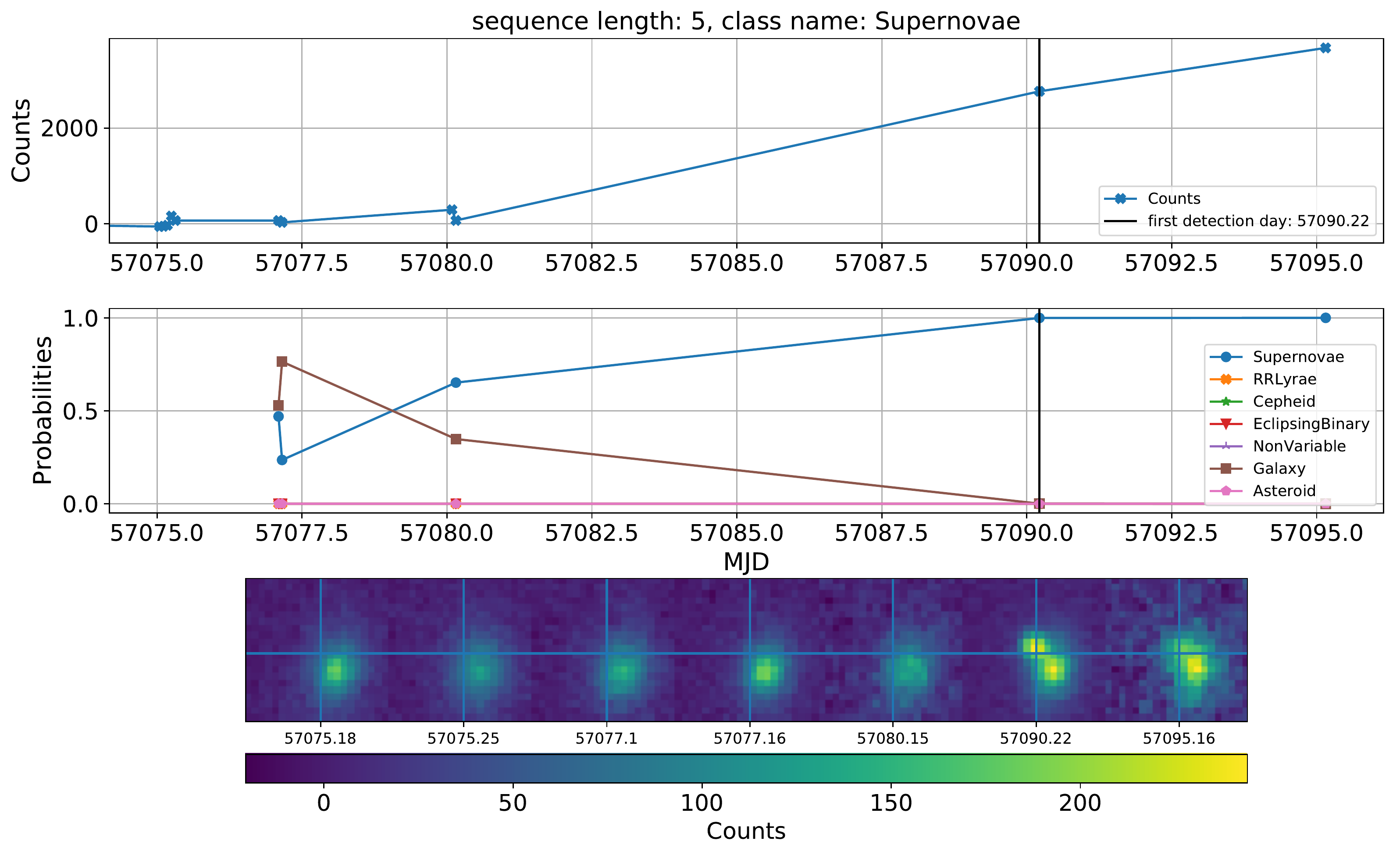}}
\caption{Example 3 of SN classification using the image sequence classifier}
\label{fig:ex3}
\end{figure}

Herein we show some examples of the image sequence RCNN classifier  working on HiTS supernovae examples. From Figures \ref{fig:ex1} to \ref{fig:ex3}, we show the light curve in counts of a supernova and the first detection time (upper plot), probabilities of the objects being of a certain class according to the model through time (mid plot) and the stamps corresponding to each observation date used as input (bottom plot). As mentioned in section \ref{sec:results}, the performance of the classifier gets better around the first detection of the supernovae (image number six).

\end{document}